\documentclass[prl,twocolumn,aps,showpacs,superscriptaddress]{revtex4}
\usepackage{graphicx}  % needed for figures
\usepackage{dcolumn}   % needed for some tables
\usepackage{bm}        % for math
\usepackage{amssymb}   % for math

\usepackage{amsmath}
\usepackage{braket}
\usepackage{color}

\begin{document}

\title{Viscous forces and bulk viscoelasticity near jamming}

\author{Karsten Baumgarten}
\affiliation{Delft University of Technology, Process \& Energy Laboratory, Leeghwaterstraat 39, 2628 CB Delft, The Netherlands}

\author{Brian P. Tighe}
\affiliation{Delft University of Technology, Process \& Energy Laboratory, Leeghwaterstraat 39, 2628 CB Delft, The Netherlands}

\date{\today}

\begin{abstract}
When weakly jammed packings of soft, viscous, non-Brownian spheres are probed mechanically, they respond with a complex admixture of elastic and viscous effects. While many of these effects are understood for specific, approximate models of the particles' interactions, there are a number of proposed force laws in the literature, especially for viscous interactions. We numerically measure the complex shear modulus $G^*$ of jammed packings for various viscous force laws that damp relative velocities between pairs of contacting particles or between a particle and the continuous fluid phase. We find a surprising sensitive dependence of $G^*$ on the viscous force law: the system may or may not display dynamic critical scaling, and the exponents describing how $G^*$ scales with frequency can change. We show that this sensitivity is closely linked to manner in which viscous damping couples to floppy-like, non-affine motion, which is prominent near jamming. 
\end{abstract}
\pacs{}

\maketitle

\section{Introduction}
Dense packings of soft, viscous, non-Brownian spheres are widely studied as a minimal model for emulsions, aqueous foams, and soft suspensions.\cite{durian95,tewari99,olsson07,langlois08,hatano09,tighe10c,boschan16,boschan17} When compressed, soft spheres ``jam'' into a marginally solid state with a shear modulus that grows continuously above a critical packing fraction $\phi_c \approx 0.84$ in 2D and 0.64 in 3D.\cite{ohern03} Close to the jamming point, structural and mechanical properties display features reminiscent of a critical point, including diverging time and length scales. \cite{bolton90,durian95,ohern03,silbert05,ellenbroek06,olsson07,tighe10c,tighe11,lerner14,karimi15,baumgarten17,khakalo17} Mechanically probing the system on finite time scales reveals a mixture of elastic and viscous response.\cite{durian95,hatano09,tighe11,boschan16,boschan17,khakalo17} However numerical studies typically represent particles' viscous interactions with their neighbors and/or the continuous fluid phase using approximate, computationally inexpensive force laws. Here we use simulations and theory to demonstrate that  viscoelastic properties of jammed solids are surprisingly sensitive to the form of the viscous force law. 

\begin{figure}[tb]
\centering
 \includegraphics[width = 0.8\columnwidth]{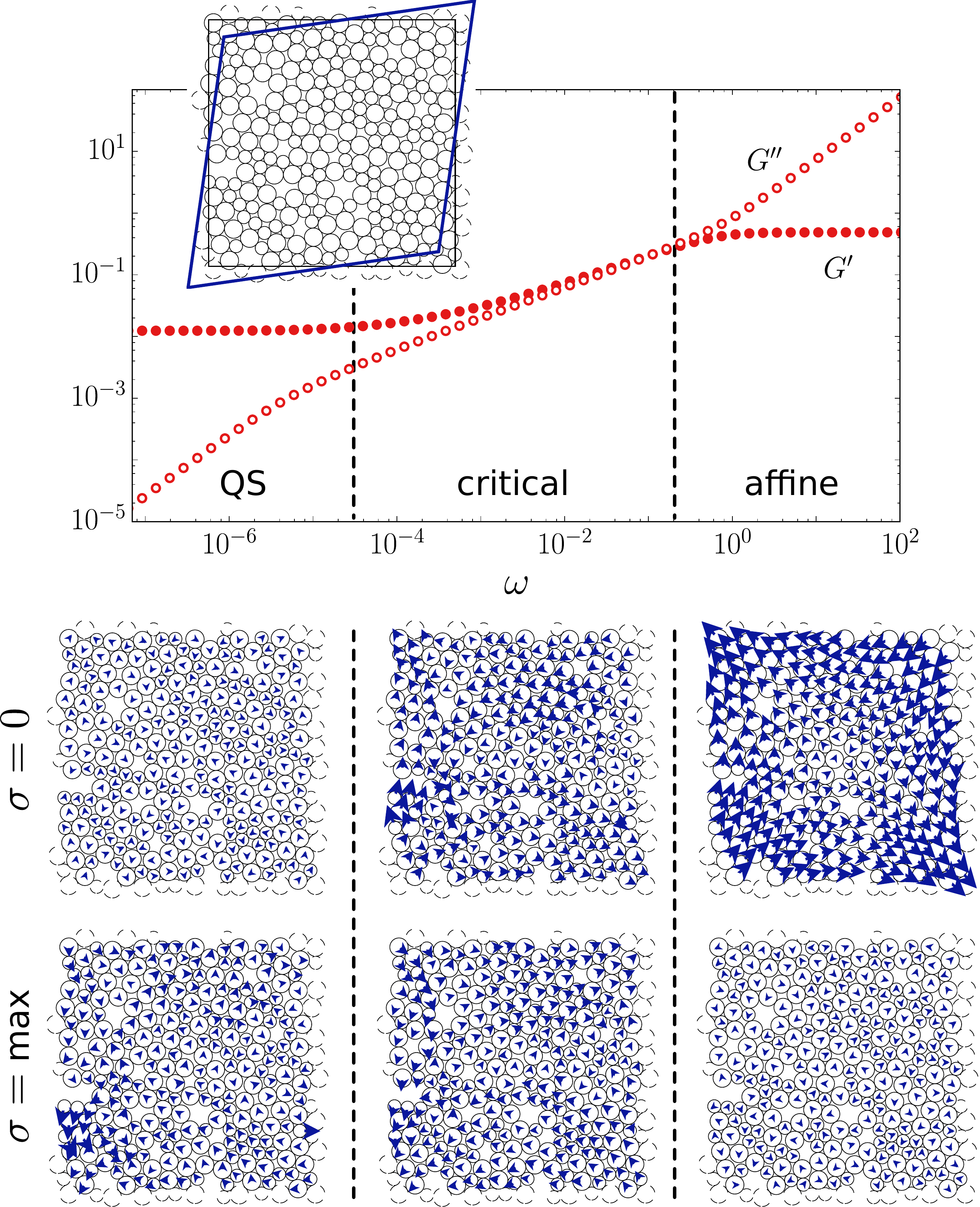}
 \caption{(top) Storage modulus $G'$ and loss modulus $G''$ of \textcolor{black}{a packing of viscous soft disks} (inset) prepared at pressure $p = 10^{-4}$ and sheared at driving frequency $\omega$. (bottom) Particle displacements evaluated at zero and peak stress amplitude for $\omega = 10^{-10}$, $10^{-3}$, and $10^4$.}
\label{fig:intro}
\end{figure}

Linear viscoelasticity is characterized by the frequency dependent complex shear modulus $G^*(\omega) = G'(\omega) + \imath \, G''(\omega)$; its real and imaginary parts are known as the storage and loss modulus, respectively, and quantify the amount of  energy stored elastically and dissipated viscously during one  cycle of oscillatory driving at angular frequency $\omega$.\cite{barnes}  The form of the complex shear modulus near jamming was first described by Tighe\cite{tighe11} for a system of soft spheres interacting via ``one-sided'' (purely repulsive) springs and linear viscous contact forces; details of the model are presented below. Characteristic features can be seen in Fig.~\ref{fig:intro}, which plots the average  $G^*$ for states prepared close to jamming. 
At both low and high frequencies, the storage modulus (filled symbols) and loss modulus (open symbols) resemble a simple \textcolor{black}{Kelvin-Voigt solid (a spring and dashpot in parallel),\cite{barnes} with} $G' \sim {\rm const}$ and $G'' \sim \omega$. There is also a critical regime at intermediate frequencies, in which both $G'$ and $G''$ scale as $\omega^{1/2}$. Similar square root scaling has been observed experimentally in foams, emulsions, and other complex fluids,\cite{liu98,cohenaddad98,gopal03,kropka10,lietorsantos11} and has been linked theoretically to strongly non-affine motion \cite{liu98}. Plots of the particles' displacements from a static initial condition, evaluated at zero and peak shear stress (Fig.~\ref{fig:intro}, bottom six panels) show that the critical regime represents a broad crossover from highly non-affine motion in the quasistatic limit at vanishing $\omega$, to strongly affine motion at high frequencies.

The square root scaling is anomalous, in the sense that simple linear interactions at the particle scale give rise to nonlinear frequency dependence in the bulk. In contrast, the frequency dependence of $G^*$ in a Kelvin-Voigt solid is consistent with a direct extrapolation from the elastic forces (linear in the particle displacements) and viscous forces (linear in the velocities). Moreover, in soft spheres the critical regime broadens on approach to the jamming transition, with its lower bound approaching zero as the confining pressure $p$ goes to zero and the system unjams.\cite{tighe11} 
This strongly suggests that critical effects lie at the origin of the square root scaling near jamming.

While spring-like forces are standard in numerical models of foams and emulsions,\cite{durian95,ohern03,olsson07,langlois08,hatano09,tighe10c,tighe11,boschan16,srivastava17} a number of alternate  proposals for viscous interactions can be found in the literature.\cite{bretherton61,durian95,denkov08,seth11,andreotti12} This variety is largely due to authors' efforts to strike a balance between physical accuracy and computational complexity. What influence does the viscous force law have on bulk viscoelastic response near jamming? In  equilibrium systems near a critical point, growing correlations wash out  particle-scale details, so that similar scaling in bulk properties can be found for different interparticle interactions.\cite{hohenberghalperin} Here we show that  the nonequilibrium jamming transition is different: the complex shear modulus near jamming is surprisingly sensitive to the form of the viscous force law. Seemingly similar choices can alter the apparent scaling exponents or eliminate dynamic critical scaling entirely. Still others lead to subtler changes in the form of correlation functions.

To probe the role of viscous damping in viscoelasticity near jamming, we implement computer simulations of Durian's bubble model, a widely studied numerical model for foams and emulsions near $\phi_c$. We investigate linear contact damping for varying ratios of the drag coefficients for normal and transverse motion, Stokes-like drag laws, and finally nonlinear damping of the relative velocities. One of our main conclusions will be to relate floppy-like, non-affine motion in the quasistatic limit to the form of the storage and loss moduli at finite frequency.  We further study the role of two-point velocity correlations and effect of pre-stress on the dynamic viscosity.

\section{The bubble model}
Durian's bubble model treats individual bubbles as non-Brownian particles interacting via elastic and viscous forces.\cite{durian95}  The equations of motion are overdamped, so that at all times the net elastic and viscous forces on a particle $i$ balance,
\begin{equation}
\vec F^{\rm el}_i + \vec F^{\rm visc}_i  = \vec 0 \,.
\label{eqn:balance}
\end{equation} 
For contact forces $\vec f_{ij}$, the corresponding net force $\vec F_i = \sum_{j(i)} \vec f_{ij}$ can be found by summing over all particles $j$ in contact with $i$. 

We consider ensembles of packings of $N$ particles in $D = 2$ spatial dimensions prepared at a target pressure $p$. $N = 32768$ unless indicated otherwise. Initial conditions are generated by minimizing the total elastic potential  energy using a nonlinear conjugate gradient algorithm, starting from particle positions placed randomly via a Poisson point process. As is typical in studies of jamming \cite{koeze16}, the packings are bidisperse to avoid crystallization, with equal numbers of large and small particles and a radius ratio 1.4:1. The systems are bi-periodic, and shear is imposed via Lees-Edwards boundary conditions. 

Units are set by the mean particle size $d$, the particle stiffness $k$, and a microscopic time scale $\tau_1$ (the latter two being introduced below). In simulations all three are set to one. However, in some cases we include the microscopic time scale in scaling relations in order to emphasize the dimensionful or dimensionless character of a relation.

\textcolor{black}{All our simulations are performed in $D = 2$ spatial dimensions, which is the upper critical dimension for the jamming transition.\cite{goodrich12} We therefore expect  the critical behavior we describe here, and in particular the values of critical exponents, to remain unchanged for $D > 2$. }

\subsection{Elastic interactions}
Elastic forces are modeled via ``one-sided springs,'' i.e.~a harmonic repulsion that acts only when particles overlap. Linear springs are a widely accepted\cite{vanhecke10,liu10a} approximate\cite{morse93,lacasse96,tighechapter,hutzler14,hohler17,winkelmann17}  description of the elastic repulsion that arises due to surface tension when spherical bubbles or droplets are deformed. The elastic force  on particle $i$ due to particle $j$ is 
\begin{equation}
\vec f_{ij}^{\, \rm el} =  \left\{\begin{array}{crl} 
-k \, \delta_{ij} \, \hat n_{ij} & {\rm for} & \delta_{ij} \ge 0 \\ 
\vec 0 & {\rm for} & \delta_{ij} < 0 \,.
 \end{array} \right. 
 \label{eqn:elasticforce}
\end{equation}
Here we have introduced the contact stiffness $k$, the overlap $\delta_{ij} = \rho_i + \rho_j - \Delta r_{ij}$, and the normal vector $\hat n_{ij} = (\vec r_j - \vec r_i)/\Delta r_{ij}$. The latter two quantities are defined in terms of the particle radii $\rho_i$ and $\rho_j$, center positions $\vec r_i$ and $\vec r_j$, and center-to-center distance $\Delta r_{ij} = |\vec r_i - \vec r_j|$. The contact stiffness $k$ is proportional to the surface tension and encodes the energetic cost of deforming a particle and thereby increasing its surface area. 

For later convenience we note that the elastic energy corresponding to Eq.~(\ref{eqn:elasticforce}) is $U = \sum_{\langle ij \rangle} U_{ij}$, where
\begin{equation}
U_{ij} =  \left\{\begin{array}{cl} 
\frac{1}{2}k \delta_{ij}^2 & \delta_{ij} \ge 0 \\ 
 0 & \delta_{ij} < 0 \,.
 \end{array} \right. 
 \label{eqn:elasticpotential}
\end{equation}
The energy change $\Delta U$ due to small perturbations away from an initial condition in mechanical equilibrium is
\begin{equation}
 \Delta U  \approx \sigma_0 \gamma V + \frac{1}{2} \sum_{\langle ij \rangle} \left[ k \left( \Delta u_{ij}^{\parallel} \right)^2 - \frac{f_{ij}^0}{\Delta r_{ij}^0} \left( \Delta u_{ij}^{\perp} \right)^2 \right] \, ,
 \label{eqn:U}
\end{equation}
where $\Delta \vec u_{ij} = \vec u_j - \vec u_i$ is the relative displacement vector, $\Delta \vec u_{ij}^\parallel = (\Delta \vec u_{ij} \cdot \hat n_{ij}) \,  \hat n_{ij}$ is its component along $\hat n_{ij}$ \textcolor{black}{, $\Delta \vec u_{ij}^\perp = \Delta \vec u_{ij} - \Delta \vec u^\parallel_{ij}$ is the transverse component, $\gamma$ the shear strain, and $V$ the volume of the packing (area in 2D). 
The shear stress $\sigma_0$ in the reference state is small with a mean value equal to zero, because the preparation protocol is isotropic. $f_{ij}^0$ and $\Delta r_{ij}^0$ are the contact force and center-to-center distance in the reference packing, respectively. The term proportional to $f_{ij}^0$ captures the influence of stress in the reference packing, i.e.~the confining pressure $p$. It is referred to as pre-stress, to distinguish it from stresses induced by the shear deformation. At several points below we present data calculated ``without pre-stress,'' which is achieved by setting $f_{ij}^0$ to zero in Eq.~(\ref{eqn:U}). This is equivalent to replacing the packing with a network of springs, each with a rest length equal to $\Delta r_{ij}^0$ from the corresponding contact.}

\begin{figure}[tb]
\centering
 \includegraphics[width = 1.0\columnwidth]{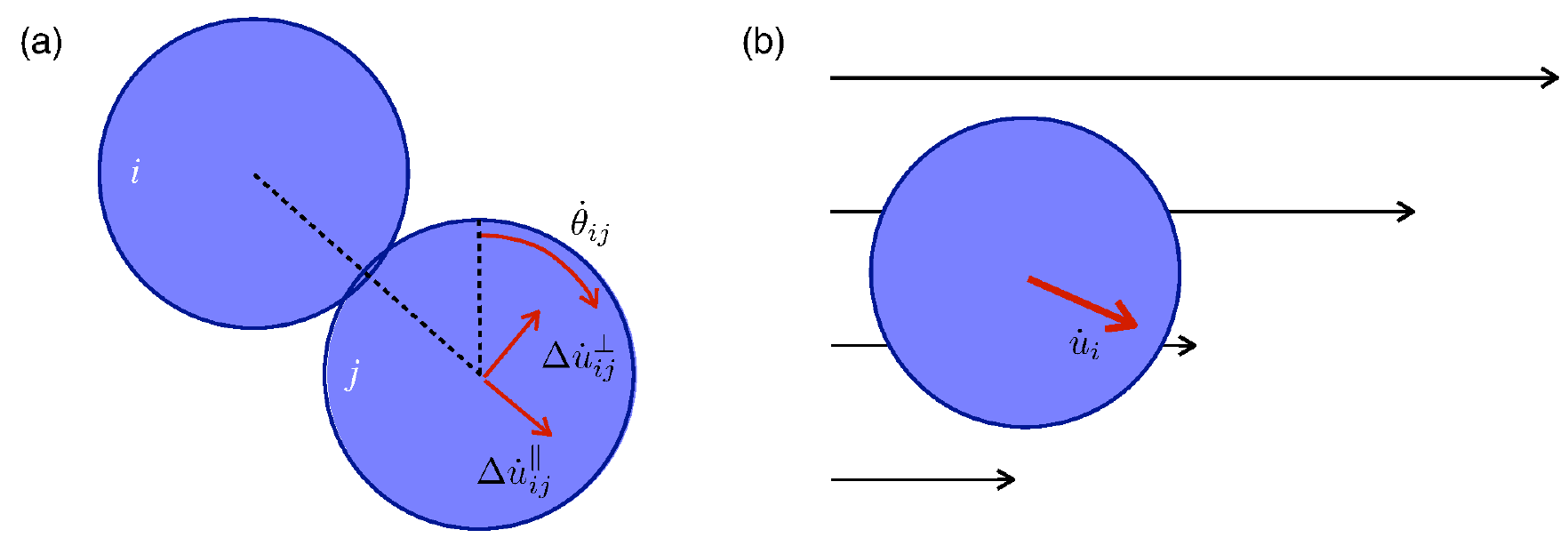}
 \caption{(a) Relative velocities in the rest frame of particle $i$. (b) Motion with respect to an affine flow field.
}
 \label{fig:sketch}
\end{figure}

\subsection{Viscous interactions}

Here we describe the several viscous force laws considered below. These can be divided in three classes: linear contact forces, linear body forces, and nonlinear contact forces.

\subsubsection{Linear contact damping}

We will explore a class of linear viscous contact force laws that damp relative velocities at the point of contact,
\begin{equation}
\vec f_{ij}^{\, \rm visc} = - k\tau_1 \left[  \Delta \dot {\vec  u}_{ij}^{\, \parallel} 
 + \beta  \, \Delta \dot {\vec u}_{ij}^{\perp,c} \right] \,.
 \label{eqn:linvisc}
\end{equation}
See Fig.~\ref{fig:sketch}a for an illustration. The quantity $\Delta \dot {\vec u}_{ij}^{\perp,c} = (\Delta \dot {u}_{ij}^\perp -  \rho_i \dot \theta_i - \rho_j \dot \theta_j)\, (\hat n_{ij} \times \hat z) $ is the tangential velocity at the contact and $\hat z$ is the out-of-plane unit vector. $\theta_i$ is the angular displacement of particle $i$ from its orientation in the initial condition. Dots indicate differentiation with respect to time. The coefficient $k\tau_1$ controls the damping of relative normal motions. It is defined in terms of a microscopic time scale $\tau_1$, which describes the exponential relaxation of two overlapping disks and sets the natural unit of time. The damping coefficient for relative transverse motion $\beta k \tau_1$ is defined by its ratio  $\beta$ to the damping coefficient for normal motion.  

The case $\beta = 1$ describes equal damping of normal and transverse motion. For brevity we refer to this case as ``balanced'' contact damping.  
Examples of prior studies employing balanced contact damping include Refs.~\cite{tewari99,langlois08,maloney08,tighe10c,tighe11,boschan16,boschan17,vagberg17}. Note that some of these studies apply damping to the relative motion of the particles' centers, neglecting particle rotations. We include rotations, as this seems more physical -- however, we have also implemented balanced damping without rotations and find the form of $G^*$ qualitatively unchanged from the results presented below.

We  also separately consider the case $\beta = 0$, in which transverse motion goes undamped. This is not a physically realistic scenario for densely packed foams and emulsions. Nevertheless, this damping law is found in the literature, presumably because it exerts no torque, eliminating the need to keep track of rotational degrees of freedom.\cite{peyneau08, hatano08, otsuki09, heussinger10} \textcolor{black}{In dilute systems with volume fractions outside the range considered here, this same force law is also a means to implement inelastic collisions.}

Finally, we also treat the case of arbitrary $\beta$. We are not aware of any prior work that has systematically varied this coefficient.

Again for later convenience, we note that the Rayleigh dissipation function corresponding to Eq.~(\ref{eqn:linvisc}) is 
\begin{equation}
 R = \frac{1}{2} k \tau_1 
 \sum_{\langle ij \rangle} \left[ 
 \left( \Delta \dot u_{ij}^{\parallel} \right)^2 
 + \beta \left( \Delta \dot u_{ij}^{\perp,c} \right)^2 
 \right] \,.
\end{equation}
\textcolor{black}{The Rayleigh dissipation function is used to implement linear damping forces in a Lagrangian formalism. Just as conservative forces are proportional to gradients of the potential energy, dissipative forces are proportional to gradients of the dissipation function.}

\subsubsection{Stokes-like drag forces}

In addition to linear contact drag, we also consider a class of linear viscous force laws in which drag enters as a body force reminiscent of Stokes drag.\cite{evansmorriss,durian95}  These can be motivated in two ways. 

In the first interpretation, drag between particles is neglected entirely. Instead drag is assumed to result from the motion of individual particles with respect to the continuous fluid phase, which itself is assumed to flow with an affine velocity profile $\vec v_{\rm aff}(\vec x) = \dot \gamma y \, \hat x$ set by the shear rate $\dot \gamma$. A particle at position $\vec r_i$ then experiences a drag force proportional to the difference between its velocity $\vec v_i$ and the affine profile (see Fig.~\ref{fig:sketch}b),
\begin{equation}
\vec F_i^{\, \rm visc} = - k\tau_1 \, [\dot {\vec  u}_i - \vec v_{\rm aff}(\vec r_i) ] \,.
\label{eqn:MF}
\end{equation}
In this interpretation, the damping coefficient $k \tau_1$ should be proportional to the fluid viscosity $\eta_F$, as specified in Stokes' law.
The dissipation function is 
\begin{equation}
R = \frac{1}{2}k \tau_1 \sum_i \left[ (\dot u_{i,x} - r^0_{i,y} \dot \gamma)^2 + \dot u_{i,y}^2 \right] + \frac{1}{2}\eta_{F} \dot \gamma^2 V \,.
\label{eqn:RMF}
\end{equation}
The second term accounts for dissipation due to shearing of the continuous fluid phase. 

An alternative interpretation of Eq.~(\ref{eqn:MF}) known as ``mean field drag'' was introduced by Durian.\cite{durian95} In this view the body force is an approximation to balanced contact damping. One assumes that the velocity of each contacting particle $j$  can be replaced with its average value at that position, which coincides with the affine velocity field. Angular velocities are set to zero. The resulting viscous force law and dissipation function are identical to Eqs.~(\ref{eqn:MF}) and (\ref{eqn:RMF}), with the caveat that $k \tau_1$ no longer has a fixed proportion to the fluid viscosity. Retaining the fluid viscosity term in the dissipation function is advisable, however, as otherwise the system could deform affinely without dissipating energy. 

Regardless of how the Stokes-like drag force is motivated, its advantage is again computational. As the equations of motion in the bubble model are overdamped, they are first order linear differential equations. Generally, these must be solved using matrix inversion (see below). However in the special case of Eq.~(\ref{eqn:MF}), the relevant inversion can be performed by hand. Prior studies using Stokes or mean field drag include Refs.~\cite{durian95,tewari99,olsson07,hatano09,andreotti12,ikeda12,tighe12,lerner12,vagberg17,khakalo17}

\subsubsection{Nonlinear contact forces}

The viscous contact force law of Eq.~\ref{eqn:linvisc} is linear in the particle velocities. However, viscous friction laws in real foams are actually nonlinear in the relative velocities. There are two classes of interactions, associated with so-called mobile and immobile surfactants, which give rise to different flow profiles within the thin films of the flow, and therefore dissipate energy differently. The case of  immobile surfactants was treated by Bretherton,\cite{bretherton61} whose drag law proportional to the $2/3$ power of velocity was subsequently verified experimentally.\cite{katgert08} More recently, Denkov and co-workers have argued for an exponent $1/2$ in the case of mobile surfactants.\cite{denkov08} Seth et al.\cite{seth11} have also suggested a nonlinear force law with exponent $1/2$ to account for elastohydrodynamic interactions between deformable particles in soft glassy matter. We therefore consider force laws of the form
\begin{equation}
\vec f_{ij}^{\, \rm visc} = - k  \tau_1 \, \left( \frac{\Delta v_{ij}^c}{\rho_0/\tau_1}\right)^{\alpha-1} \, \Delta{\vec v_{ij}^{\, c}}\,,
\label{eqn:nonlin}
\end{equation}
where $\Delta \vec v_{ij}^{\, c} = \Delta \dot{\vec u}_{ij}^{\parallel} + \Delta \dot{\vec u}_{ij}^{\perp,c}$ is the relative velocity at the contact.
The constant $\rho_0$ has units of length and is required for dimensional consistency. We set it to 1. 
We will consider a range of exponents $\alpha$, including the physically relevant values of $2/3$ and $1/2$.

\subsection{Equations of motion}

To solve for the complex shear modulus, it is useful to rewrite the equations of motion, Eq.~(\ref{eqn:balance}), in matrix form. Following Ref.~\cite{tighe11}, the equations of motion can be expressed as
\begin{equation}
\label{eq:dgl}
 \hat{\mathcal{K}} \ket{Q(t)} + \hat{\mathcal{B}} \ket{\dot{Q}(t)} = \ket{F(t)} \,.
\end{equation}
The Hessian matrix $\hat{\mathcal{K}}$ and the damping matrix $\hat{\mathcal{B}}$ are defined in terms of the elastic potential energy $U$ and the Rayleigh dissipation function $R$, 
\begin{equation}
 \mathcal K_{mn} = \left. \frac{\partial^2 U}{\partial Q_m \partial Q_n} \right|_{\ket Q = \ket 0} \quad \mathcal B_{mn} = \left. \frac{\partial^2 R}{\partial \dot Q_m \partial \dot Q_n} \right|_{\ket{\dot{Q}} = \ket 0} \, .
 \label{eqn:eom}
\end{equation}
The $3N+1$-component vector $\ket{Q} = (u_{1x}, u_{1y}, \ldots, \theta_1, \theta_2, \ldots, \gamma)$ contains all degrees of freedom, including the amplitude $\gamma$ of the pure shear strain experienced by the box. The reference packing is defined as the state $|Q\rangle = |0\rangle$. The vector $\ket{F}$ contains the generalized forces conjugate to each of the components of $\ket{Q}$. The component conjugate to $\gamma$ is equal to \textcolor{black}{$\delta \sigma \,V = (\sigma - \sigma_0)V$, where $\sigma$ is the shear stress.} 

The Fourier transform of Eq.~(\ref{eqn:eom}) gives
\begin{equation}
 \left(\hat{\mathcal{K}} + \imath \omega \hat{\mathcal{B}} \right) \ket{Q^*(\omega)} =  \delta \sigma \, V \ket{\hat \gamma}
 \label{eq:fourier_dgl}
\end{equation}
where $\omega$ is the angular frequency. Note that $|Q^*(\omega)\rangle$ is complex. We impose a generalized forcing term pointing along the $\gamma$-coordinate, i.e.~$\ket{F} \propto \ket{\hat \gamma} = (0, 0, \ldots, 1)$. All other generalized forces are zero (body forces and torques are balanced). The equations of motion are therefore reduced to a a set of complex linear equations which can be solved numerically for each frequency $\omega$.

The complex shear modulus can be determined by solving Eq.~(\ref{eq:fourier_dgl}) for the complex vector $\ket{Q^*(\omega)}$ using standard linear algebra routines. The resulting shear strain is $\gamma^*(\omega) = \langle \hat \gamma | Q^*(\omega) \rangle$. The complex shear modulus is then 
\begin{equation}
G^*(\omega) \equiv G'(\omega) + \imath G''(\omega) = \frac{\gamma^*(\omega)}{\delta \sigma} \,.
\end{equation}

\begin{figure*}[tb]
\centering
\includegraphics[width = 0.8\linewidth]{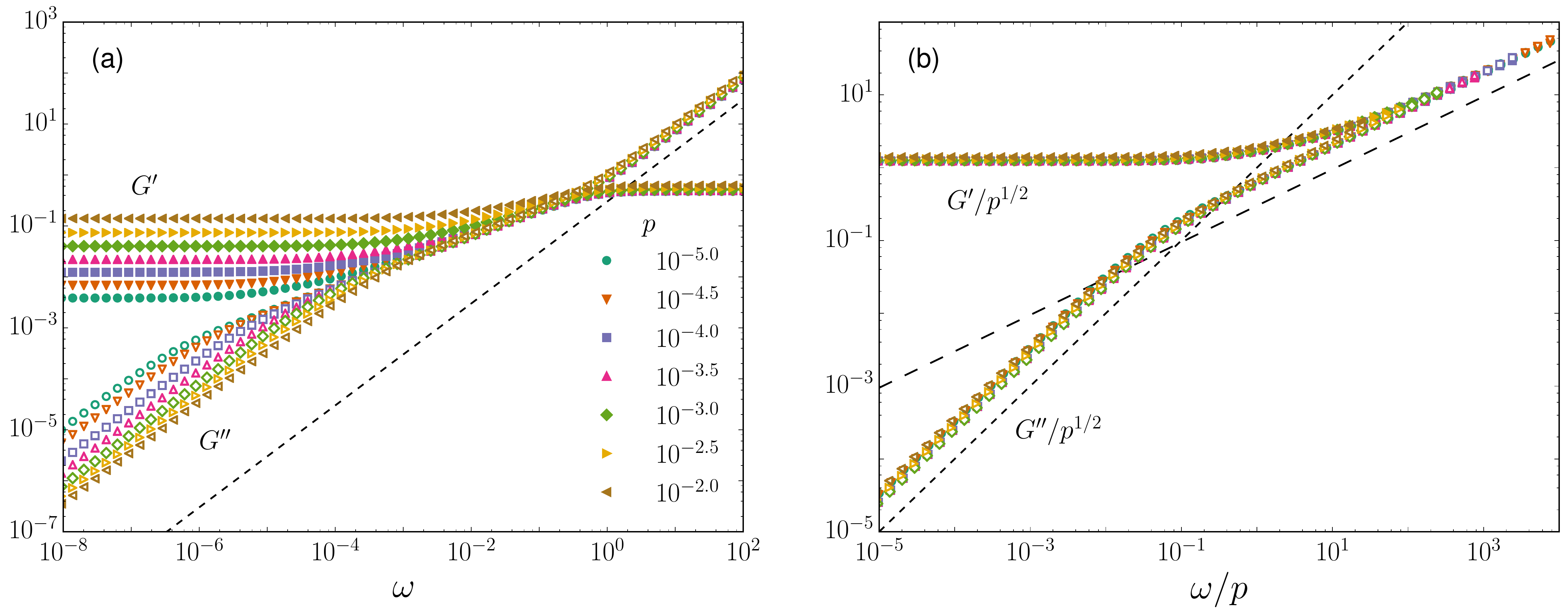} 
 \caption{(a) The storage and loss moduli, $G'$ and $G''$, for balanced contact damping ($\beta = 1$). The dashed line has slope 1.
(b) Collapse of the same data to two critical scaling functions. The short- and long-dashed lines have slopes of 1 and 1/2, respectively.}
 \label{fig:pressure_sweep}
\end{figure*}

\section{Linear contact damping}

We now consider the complex shear modulus in the presence of linear contact damping. We begin with balanced damping, i.e.~Eq.~(\ref{eqn:linvisc}) for $\beta = 1$. This scenario was already extensively studied in Ref.~\cite{tighe11}, and provides a useful point of comparison for alternative viscous force laws. Here we highlight the main results.

\subsection{Balanced damping}

Balanced linear contact damping was discussed above for the case $p = 10^{-4}$ -- see Fig.~\ref{fig:intro}. 
We can gain further insight by varying the distance to jamming. In Fig.~\ref{fig:pressure_sweep}a we plot the complex shear modulus as a function of frequency for a range of pressures $p = 10^{-5} \ldots 10^{-2}$. In all cases the same quasistatic, critical, and affine regimes identified in Fig.~\ref{fig:intro} are evident. However the crossover frequency $\omega^* \equiv 1/\tau^*$ from the quasistatic to the critical regime shifts to lower values as $p \rightarrow 0$, indicating that the time scale $\tau^*$ diverges at the jamming point. The crossover from critical to high frequencies, on the other hand, is insensitive to pressure; it occurs for $\omega \sim {\cal O}(1)$ in all cases. We can infer that the quasistatic and critical regimes are intimately related to the jamming transition, while the high frequency response does not have a critical character. 

Inspired by the above observation, we now restrict our focus to frequencies $\omega < 1$. A more rigorous derivation of the following results is found in Ref.~\cite{tighe11}. Our approach here is more heuristic and begins with the scaling ansatz
\begin{equation}
\frac{G^*}{G_0}   =  \mathcal{G}^* \left( {\omega}{\tau^*} \right) \,\,\, {\rm for} \,\,\,\, \omega < {\cal O}(1) \,,
\label{eqn:ansatz}
\end{equation}
which relates the dimensionless ratio $G^*/G_0$ to the dimensionless product $\omega \tau^*$. As discussed below, the quasistatic shear modulus scales as $G_0  \sim p^\mu$ with $\mu = 1/2$. Similarly, we assume that $\tau^*$ diverges at the jamming point, 
\begin{equation}
{\tau^*} \sim \frac{1}{p^\lambda} 
\end{equation}
for some positive exponent $\lambda$.
The real and imaginary parts of the scaling function ${\cal G}^* = {\cal G}'  + \imath {\cal G}''$ satisfy\begin{align}
{\cal G}'(x) \sim \left \lbrace
	\begin{array}{cl}
	1 & x < 1 \\
	x^\Delta & x > 1
	\end{array} \right.
\label{eqn:calGp}
\end{align}
and
\begin{align}
{\cal G}''(x) \sim \left \lbrace
	\begin{array}{cl}
	x & x < 1 \\
	x^\Delta & x > 1\,.
	\end{array} \right.
\label{eqn:calGpp}
\end{align}
The forms ${\cal G}' \sim 1$ and ${\cal G}'' \sim x$ for small $x$ are the simplest choices respecting the symmetry properties of the storage and loss moduli, which are even and odd functions, respectively.  The power laws ${\cal G}' \sim x^\Delta$ and ${\cal G}'' \sim x^\Delta$ represent non-trivial assumptions.  The same exponent $\Delta$ must appear in both the real and imaginary parts to satisfy the Kramers-Kronig relations. 

The scaling ansatz of Eqs.~(\ref{eqn:ansatz}-\ref{eqn:calGpp}) is tested in Fig.~\ref{fig:pressure_sweep}b, which plots $G'/p^\mu$ and $G''/p^\mu$ with $\mu = 1/2$ versus $\omega/p^\lambda$ with $\lambda = 1$. The resulting collapse is excellent. As expected, the real and imaginary parts of the scaling function are constant and linear, respectively, for low values of the rescaled frequency. There is a crossover around $\omega/p \sim {\cal O}(1)$ to a power law with exponent $\Delta \approx 0.5$ (long dashed line). This is the $\omega^{1/2}$ scaling discussed above. 

The scaling collapse in Fig.~\ref{fig:pressure_sweep} empirically determines the values of the critical exponents; they are $\mu = 1/2$, $\lambda = 1$, and $\Delta = 1/2$. The value of $\mu$ is fixed by the known scaling of $G_0$.  The exponent $\Delta$ is related to $\mu$ and $\lambda$. To see this, note that one generally expects the moduli to remain finite except possibly at the critical point, where both $p$ and $\omega$ go to zero. In the case where $p = 0$ and $\omega > 0$, Eqs.~(\ref{eqn:ansatz}-\ref{eqn:calGpp}) predict that both moduli scale as $p^{\mu - \lambda \Delta} \omega^\Delta$, which remains finite only if $\Delta = \mu/\lambda = 1/2$. It remains to motivate  $\lambda = 1$, which we do in Section \ref{sec:floppy}.

\subsection{``Imbalanced'' contact damping ($\beta \neq 1$)}

In this section we probe the effects of undamped sliding motion, with emphasis on the limit $\beta = 0$. Our main result is to show that imbalanced damping ``kills'' dynamic critical scaling near jamming. 

\begin{figure}[tb]
 \centering
  \includegraphics[width = 0.8\columnwidth]{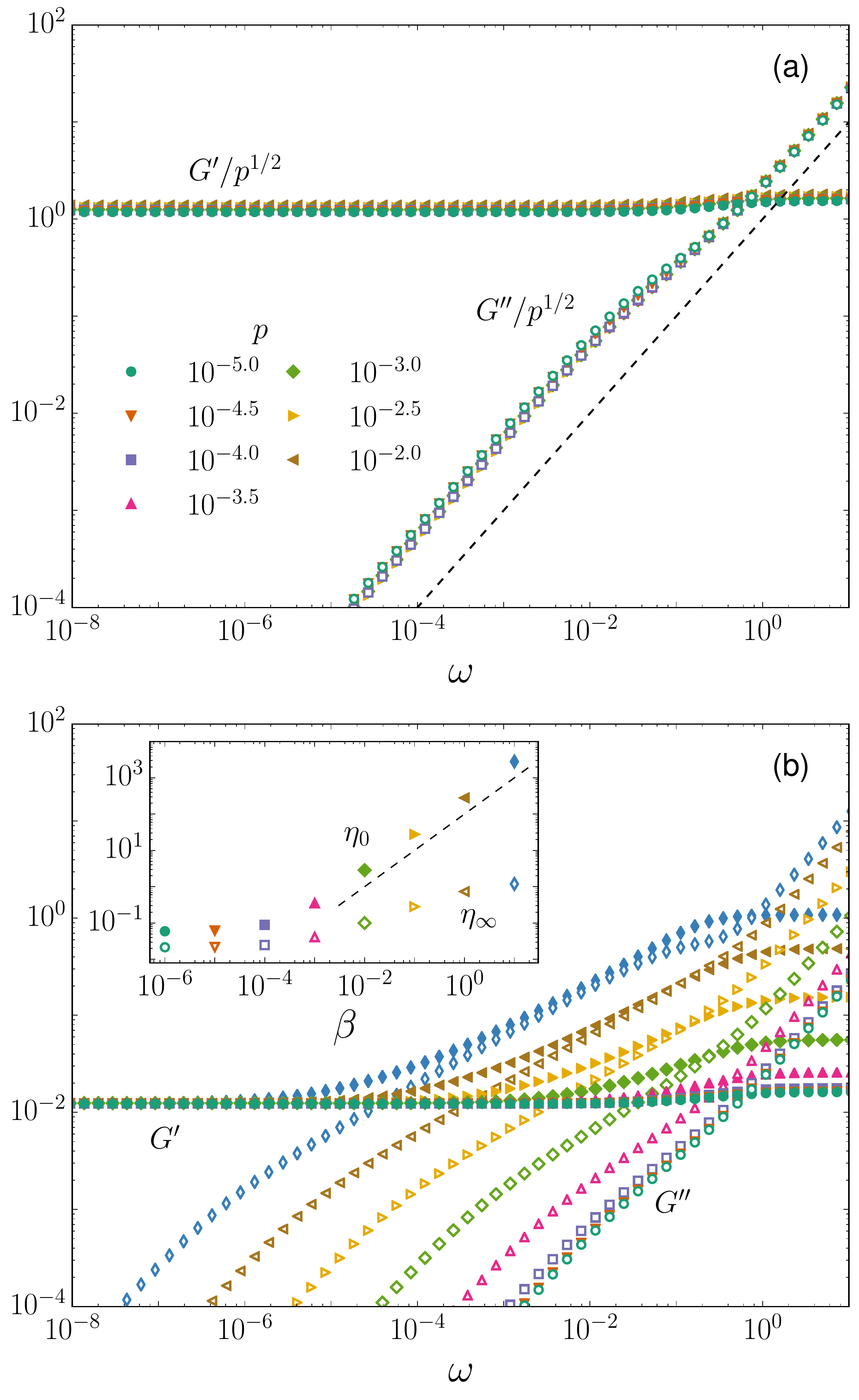}
 \caption{\textcolor{black}{(a) Storage and loss modulus for a system without transverse damping ($\beta = 0$). (b) $G^{\prime}$ and $G^{\prime\prime}$} for systems at pressure $p = 10^{-4}$ and varying transverse damping $\beta$. (inset) Dynamic viscosity $\eta_0$ and affine viscosity $\eta_\infty$ for the same data, denoting the low and high frequency limits of $G''/\omega$. \textcolor{black}{In both figures the dashed line has slope 1.}}
 \label{fig:B_sweep}
\end{figure}

It is useful  first to consider response in the absence of the pre-stress term, i.e.~by setting $f_{ij}^0 = 0$ in Eq.~(\ref{eqn:U}). The Hessian and damping matrices are then directly proportional, $\hat{\mathcal{K}} = \tau_1 \hat{\mathcal{B}}$, allowing  Eq.~\eqref{eq:fourier_dgl} to be solved exactly in terms of $G_0$,
\begin{equation} 
G^* = G_0 (1 + \imath \tau_0 \omega) \,.
\label{eqn:KV}
\end{equation}
The resulting complex shear modulus is that of a Kelvin-Voigt  element, the simplest viscoelastic solid -- the storage modulus is flat, while the loss modulus is linear over the entire range of $\omega$. Re-introducing the pre-stress term breaks the direct proportionality between $\hat{\cal K}$ and $\hat{\cal B}$, but produces only mild changes in the moduli, as shown in Fig.~\ref{fig:B_sweep}a (open and filled squares). Moreover, data for a range of pressures close to the jamming point can all be collapsed by rescaling the storage and loss moduli by $p^{0.5}$. Note that the frequency axis does not need to be rescaled, indicating the absence of a diverging time scale.

We emphasize that a seemingly simple change to the viscous force law, namely setting the damping coefficient for sliding motion to zero, has produced a dramatic and qualitative shift in the viscoelastic response. More precisely, the intermediate regime, identified above when $\beta = 1$, has completely vanished. Recall that this regime is a manifestation of dynamic critical scaling and dominates the response for a wide range of frequencies near jamming. In this sense setting $\beta = 0$ kills dynamic critical scaling.

What happens for intermediate values of $\beta$?  In Fig.~\ref{fig:B_sweep}b we plot $G^*$ for fixed $p$ and a range of $\beta$ over seven decades.  One sees that the critical regime gradually appears, and for sufficiently large $\beta$ the moduli resemble their form for $\beta = 1$. This suggests that it is reasonable to speak of weakly and strongly damped sliding motion. We quantify this distinction more precisely below.  

\subsection{Relation to floppiness in quasistatic response}
\label{sec:floppy}

The dynamic critical scaling of Eq.~(\ref{eqn:ansatz}), and the critical exponent $\lambda$ in particular, can be related to the scaling relations for normal, transverse, and non-affine motion in quasistatic response. This link is motivated by the observation that for asymptotically low driving frequencies, the particles' trajectories must approach their quasistatic ($\omega \rightarrow 0$) form.

Packings at the jamming point are isostatic, meaning they have just enough contacts to constrain all particle motions (except for a few individual ``rattlers'', which can be removed from the analysis). Consider breaking a contact in a packing at the jamming transition, where all contacting particles are ``kissing'' and $f_{ij}^0 = 0$. The broken contact removes a constraint and therefore introduces a floppy mode, an infinitesimal motion of the particles that can be performed without work. By considering the energy expansion of Eq.~(\ref{eqn:U}), one sees that all relative normal motions in a floppy mode must be zero -- floppy motions are sliding motions, in which all relative motion between particles is transverse to the contact. Jammed packings do not have floppy modes, but the eigenmodes of the Hessian remain ``floppy-like,'' i.e.~transverse/sliding motion dominates.\cite{ellenbroek06,ellenbroek09} This feature is also found in the response to shear, which is dominated by low frequency modes \cite{tighe11}. Through careful analysis of the modes, it is possible to show that the shear modulus scales as $G_0 \sim p^{1/2}$.\cite{wyartannales,tighe11} Here we take this scaling relation as a given and, following Ref.~\cite{ellenbroek09}, infer its consequences for the typical relative normal and transverse displacement amplitudes, $\Delta u^\parallel$ and $\Delta u^\perp$, as well as the typical amplitude of non-affine displacements $u_{\rm na}$. 

By definition, the change in elastic energy $\Delta U \equiv U - U_0$ due to an infinitesimal shear strain $\gamma$ is $\Delta U  = (1/2)G_0V \gamma^2$. Momentarily neglecting the pre-stress term in Eq.~(\ref{eqn:U}), which should be small as $p \rightarrow 0$, we anticipate that the typical relative normal motion scales as $(\Delta u^\parallel)^2 \sim G_0 \gamma^2$, or 
\begin{equation}
\frac{\Delta u^\parallel}{\gamma} \sim p^{1/4}  \,.
\label{eqn:upar}
\end{equation}
This scaling relation is consistent with our expectation that relative normal motion vanishes at the jamming point. 
We now re-introduce the non-positive pre-stress term in Eq.~(\ref{eqn:U}) in order to determine $\Delta u^\perp$. \textcolor{black}{The first and second terms in brackets in Eq.~(\ref{eqn:U})} have typical values $(\Delta u^\parallel)^2$ and $p (\Delta u^\perp)^2$, respectively; in the latter case we have used the face that the typical force in the reference packing is proportional to the pressure. While mechanical stability requires the total energy change $\Delta U$ to be positive,\cite{dagois-bohy12} the system can minimize its deformation energy by organizing its motion to make the magnitude of the pre-stress term as large as possible -- in other words, if the bound $p (\Delta u^\perp)^2 \lesssim (\Delta u^\parallel)^2$ is saturated.  This gives 
\begin{equation}
\frac{\Delta u^\perp}{\gamma} \sim \frac{1}{p^{1/4}} \,.
\label{eqn:uperp}
\end{equation}
This relation relies on the (reasonable) assumption that the typical contact force scales linearly with the pressure. As expected, the amount of sliding motion grows dramatically and ultimately diverges as the system approaches the jamming point. 

Finally, we consider the typical amplitude of non-affine displacements $u_{\rm na}$. Bond vectors $\Delta \vec r_{ij}^{\, 0}$ in the reference packing are randomly oriented, so there is a local competition between energetically favorable sliding at the particle scale, and globally imposed affine motion. Therefore we expect  the typical non-affine amplitude to be comparable to the typical relative displacement amplitude, which is dominated by transverse motion, i.e.
\begin{equation}
\frac{u_{\rm na}}{\gamma} \sim \frac{1}{p^{1/4}} \,.
\label{eqn:una}
\end{equation}
Hence non-affine motion is the natural consequence of floppy-like motion near jamming. 

\begin{figure}
\centering
 \includegraphics[width = 0.8\columnwidth]{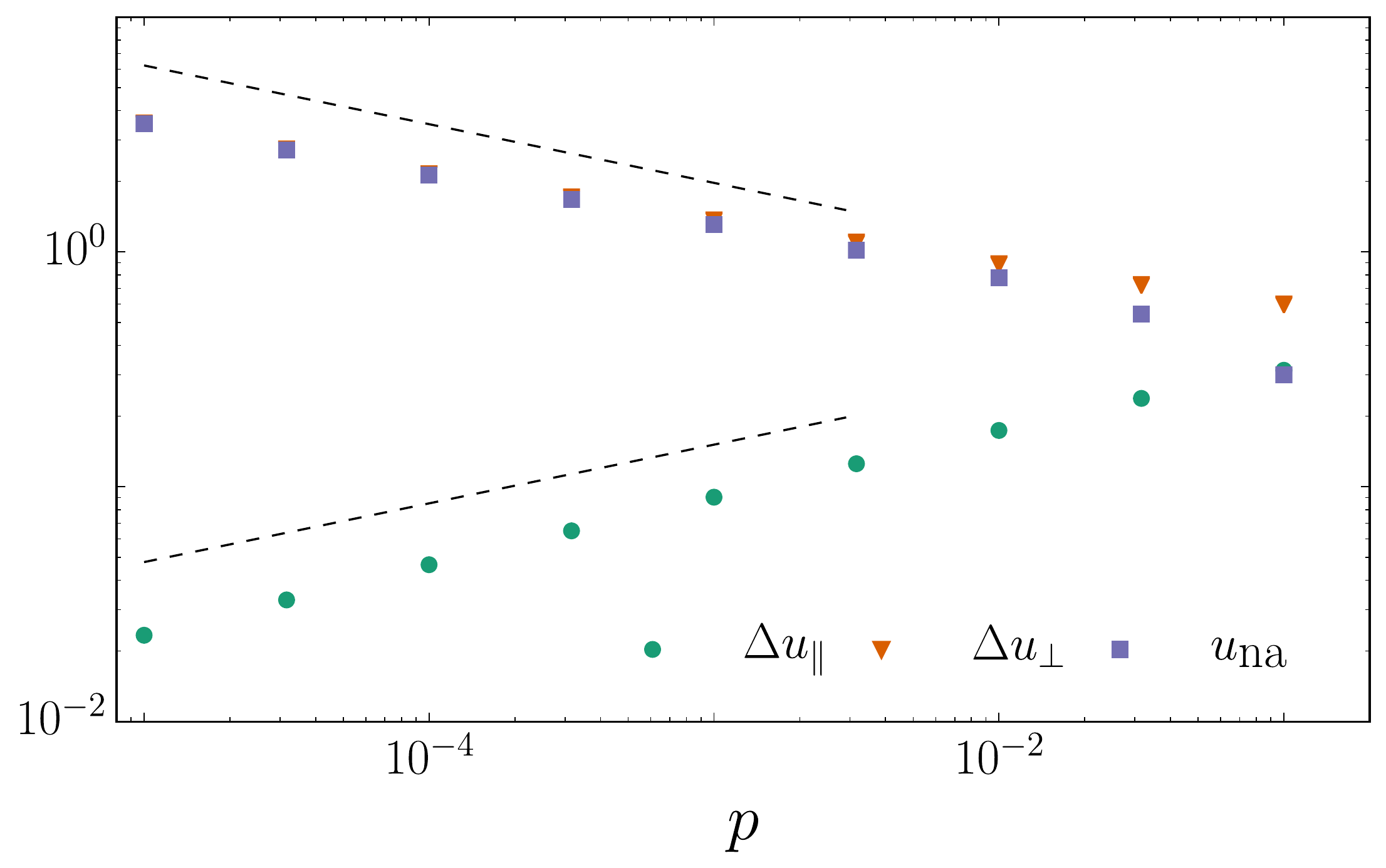}
 \caption{Scaling of the relative normal, relative transverse, and non-affine motion as a function of pressure. Dashed lines have slopes of $\pm 1/4$.}
 \label{fig:QS}
\end{figure}

Eqs.~(\ref{eqn:upar}-\ref{eqn:una}) have previously been derived and tested numerically by Ellenbroek et al.~and Wyart et al.\cite{ellenbroek09,wyart08} For completeness we verify them again in Fig.~\ref{fig:QS}, which plots the median of the probability density function of $|\Delta u_\parallel|$, $|\Delta u_\perp|$, and $|u_{\rm na}|$ for varying $p$ while neglecting the pre-stress term. Results including pre-stress show compatible trends, albeit with more noise; we revisit the role of pre-stress below. Plots of the means show the same trend for $\Delta u_\perp$ and $u_{\rm na}$, but  $\Delta u_\parallel$ develops a plateau at low $p$ due to a long tail of the PDF.

We now use the quasistatic relations (\ref{eqn:upar}-\ref{eqn:una}) to determine the critical exponent $\lambda$.
The $\omega \rightarrow 0$ limit of the dissipation function is proportional to the dynamic viscosity, $R = \eta_0 (\omega \gamma_0)^2 V/2$, where $\gamma_0$ is the maximum strain amplitude. At the same time, from the viscous force law one anticipates $R \sim \vec f^{\rm visc}\cdot \Delta {\vec v}  \sim  \omega^2[ (\Delta u^\parallel)^2 + \beta (\Delta u^\perp)^2]$. Invoking  Eqs.~(\ref{eqn:upar}) and (\ref{eqn:uperp}) gives
\begin{align}
\eta_0 &\sim 
\left(\frac{\Delta u^\parallel}{\gamma_0}\right)^2  
+ \beta \left(\frac{\Delta u^\parallel}{\gamma_0}\right)^2  \nonumber \\
&\sim (p/k)^{1/2} + \frac{\beta}{(p/k)^{1/2}} \,.
\label{eqn:eta0}
\end{align}
For balanced damping and $p \ll 1$, the second term dominates and $\eta_0 \sim 1/p^{1/2}$. Comparing to Eqs.~(\ref{eqn:ansatz}-\ref{eqn:calGpp}), which require $\eta_0 = G_0 \, \tau^* \sim p^{\mu-\lambda}$, it follows that $\lambda = 1$ and $\Delta = 1/2$. Hence we can motivate the exponents in the scaling functions (\ref{eqn:calGp}) and (\ref{eqn:calGpp}). 

Eq.~(\ref{eqn:eta0}) is compatible with our numerical results for undamped sliding ($\beta = 0$), as well. Then only the first term is present and $\eta_0 \sim p^{1/2}$ -- it vanishes rather than diverges.

One can also consider the case of arbitrary $\beta$. The second term will always dominate for sufficiently low pressure; hence the dynamic viscosity diverges for any finite damping of sliding motion. In this sense the case $\beta = 0$ is singular. For arbitrary $\beta > 0$ the  crossover frequency where the quasistatic regime ends and the linear regime begins scales as $\omega^* \sim p/\beta$. We have seen above that the critical regime ends at a frequency $\omega\tau_1 \sim  {\cal O}(1)$. Hence the critical regime, with its $\omega^{1/2}$ scaling in $G'$ and $G''$, is avoided entirely whenever $\omega^* \gg 1$, or $\beta \ll \beta^* \sim p$. This crossover is evident in Fig.~\ref{fig:B_sweep}b. The scale $\beta^*$ provides a convenient dividing line between cases of strong and weak damping of transverse motion.

\begin{figure}[tb]
\centering
\includegraphics[width = 0.8\columnwidth]{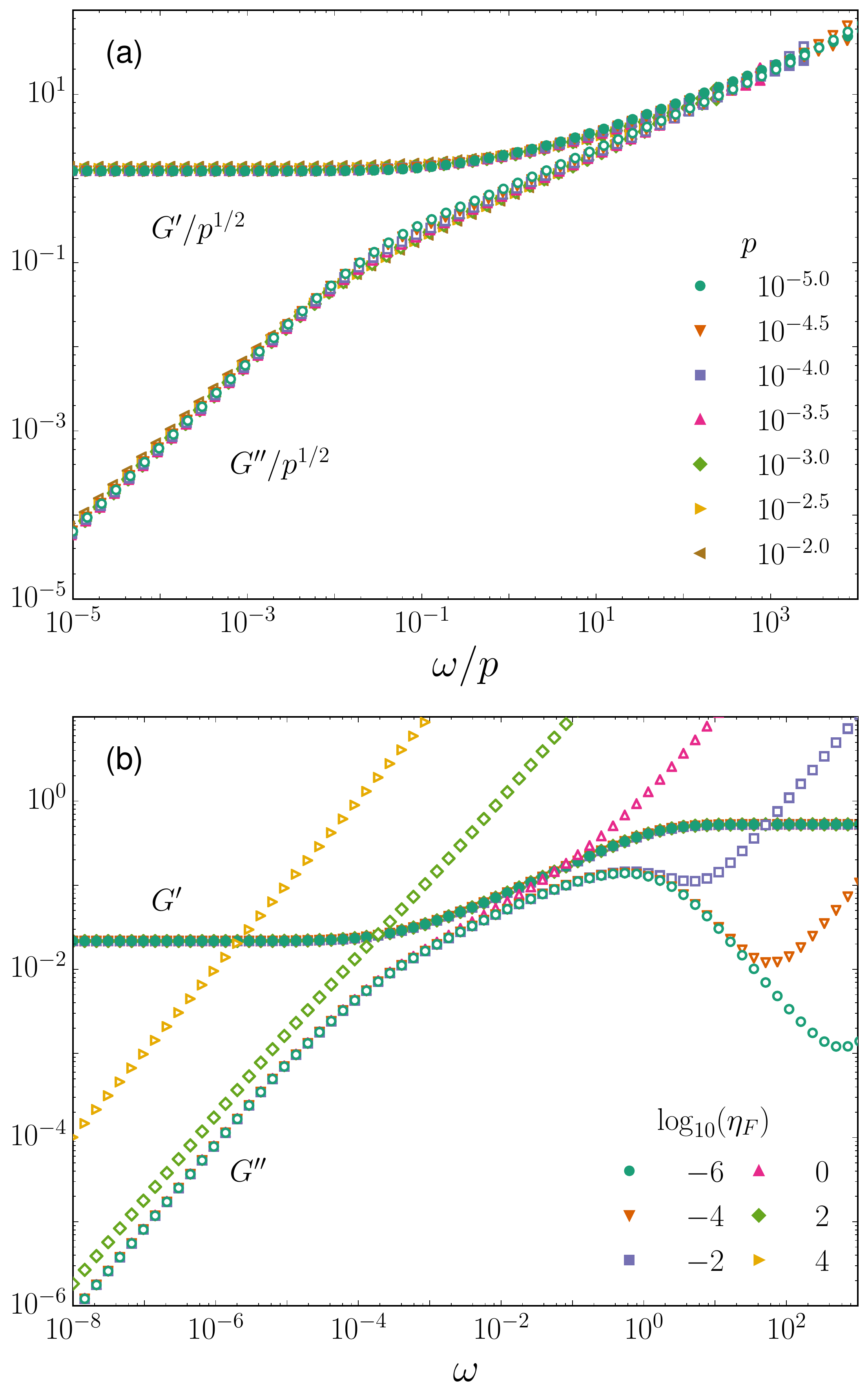}
 \caption{\textcolor{black}{(a)} Critical scaling collapse of the storage and loss moduli for Stokes drag and fluid viscosity $\eta_F = 1$.
 \textcolor{black}{(b)} Storage and loss moduli for $p = 10^{-4}$  and varying $\eta_F$.
}
 \label{fig:stokes_drag}
\end{figure}

\begin{figure*}
\centering
 \includegraphics[width = 0.87\textwidth]{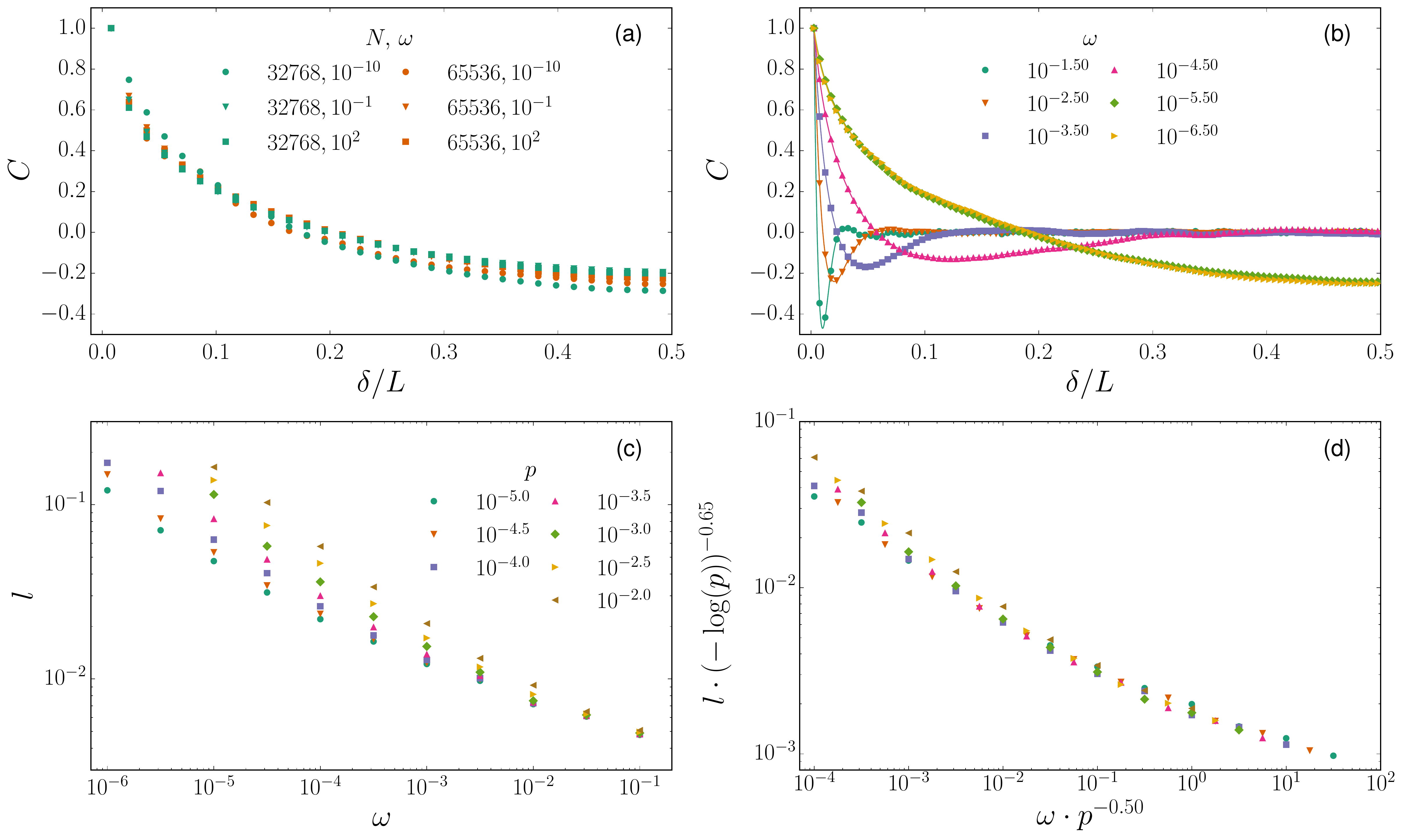}
 \caption{(a) Two-point correlation function $C$ of the transverse (hence non-affine) particle displacements with balanced contact damping. (b) The same correlation function $C$ for Stokes drag. (c) Length scale $l$ corresponding to roots of the curves in (b). (d) Data collapse of $l$.}
 \label{fig:correlations}
\end{figure*}

\section{Stokes Drag}
We now turn to the case of linear viscous body forces, i.e.~the mean field or Stokes-like drag of Eq.~(\ref{eqn:MF}). In Fig.~\ref{fig:stokes_drag}a we plot the complex shear modulus for Stokes drag for varying pressure and a fluid viscosity $\eta_F = 1$. We find dynamic critical scaling with the same critical exponents $\mu = 1/2$, $\lambda = 1$, and $\Delta = 1/2$ as for balanced contact drag. Hence it appears that Stokes drag falls into the same universality class as strongly damped relative transverse motion. 

As in the previous Section, the above result can be rationalized on the basis of quasistatic scaling relations. The key observation is that the typical non-affine motion $u_{\rm na}$ and the relative transverse motion $\Delta u^\perp$ diverge in the same way as  the pressure tends to zero; {\em cf}. Eqs.~(\ref{eqn:uperp}) and Eq.~(\ref{eqn:una}). For Stokes drag the dissipation function scales as $R \sim  (u_{\rm na} \omega)^2$, again giving $\eta_0 \sim (p/k)^{-1/2}$. 

Recall that if one considers the Stokes drag term to be a mean field approximation for balanced contact damping, then the fluid viscosity $\eta_F$ can vary independently of the damping coefficient $k\tau_1$. We probe the dependence of $G^*$ on $\eta_F$ in Fig.~\ref{fig:stokes_drag}b by varying $\eta_F$ over ten decades. We observe that the fluid viscosity contributes a linear term $\eta_F \omega$ to the loss modulus, which is always dominant at sufficiently high frequencies. For large $\eta_F$ and/or low pressures satisfying  $\eta_F \gg  1/p^{1/2}$,  the loss modulus becomes linear for all frequencies. In this event the critical properties of the loss modulus are obscured, but criticality is still apparent in the storage modulus.

\subsection{Correlations}

Despite the similarity in their viscoelastic response, we find a striking difference in the spatial correlations of non-affine displacements between the cases of  linear viscous body forces and balanced contact damping. 

For a system undergoing simple shear in the $x$-direction, correlations of the non-affine displacements between particles separated by a distance $\delta_{ij} = |x_i - x_j|$ can be quantified with the two-point correlation function $C = \langle u'_{i,y}(x_i) \, u'_{j,y}(x_i + \delta_{ij}) \rangle / \langle (u'_{i,y})^2 \rangle $.
Here ${ u}_{i,y}'$ is the $y$-component of the real part of the complex displacement vector of particle $i$ with $x$-coordinate $x_i$.  The average $\langle \cdot \rangle$ runs over all particle pairs within a narrow ``lane'', hence $C$ is a function of $|\delta_{ij}|$.   We have verified that $C$ becomes independent of the lane width for sufficiently small values. We have also confirmed that results using the imaginary part ${u}_{i,y}''$ are indistinguishable.

Non-affine correlation functions in weakly jammed solids have been studied previously for three cases. DiDonna and Lubensky\cite{didonna05} and Maloney\cite{maloney06} showed there is no characteristic length scale in quasistatic linear elastic response; instead $C$ collapses when distances are rescaled by the box size $L$. Heussinger and Barrat found compatible results for quasistatic shear flow. Olsson and Teitel\cite{olsson07} found that the same correlation function does select a growing length scale, independent of $L$, in shear flow at finite rate using Stokes drag. However Tighe et al.\cite{tighe10} showed that the form of $C$ resembles quasistatic linear response when one uses balanced contact damping instead of Stokes drag. Hence there remain important open questions about correlations at finite driving rate and the role of the viscous force law. Here we fill a gap in the literature, namely linear response at finite rates. 

In Fig.~\ref{fig:correlations}a we plot $C$ for balanced contact damping at a single pressure, two system sizes, and three values of the frequency $\omega$ separated by twelve decades. There is a monotonic decay of the correlations, with little dependence on the frequency. The shape is also independent of the pressure (not shown). The data collapse when plotted as a function of $\delta/L$.  Hence two-point displacement correlations provide no evidence of a growing length scale near jamming; snapshots of the velocities display ``swirls'' with a characteristic radius of approximately one quarter of the box size.

Correlations for Stokes-like drag display a strikingly different shape, as shown in Fig.~\ref{fig:correlations}b. $C$ possesses a minimum that shifts to larger distances with decreasing $\omega$. For the lowest plotted frequencies, $\omega = 10^{-5.5}$ and $10^{-6.5}$, the minimum is no longer clearly identifiable and the shape of $C$ begins to resemble the form for balanced contact damping.
One can define a correlation length $l$ from the point where $C$ crosses the $x$-axis, plotted in Fig.~\ref{fig:correlations}c. We find a length scale that grows with decreasing frequency, before reaching a plateau with a height of approximately $L/4$. Focusing on length scales below this plateau, we find empirically that a reasonable data collapse is achieved by plotting $l/(-\ln{p})^{0.65}$ versus $\omega/p^{0.5}$, implying that the length scale would diverge at the jamming point ($p \rightarrow 0$ and $\omega \rightarrow 0$) in thermodynamically large systems. We note that log corrections are typical in systems at their upper critical dimension, which is indeed $D = 2$ for the jamming transition.\cite{goodrich12,goodrich14}.

The takeaway is that the form of the correlation function at finite rate is strongly sensitive to the viscous force law. For balanced contact damping there is no evidence of a diverging length scale. For Stokes drag there is a growing correlation length that is cut off by the box size as the frequency is sent to zero. 

\subsection{Finite Size Effects}
\begin{figure}[tb]
\centering
 \includegraphics[width = 0.8\columnwidth]{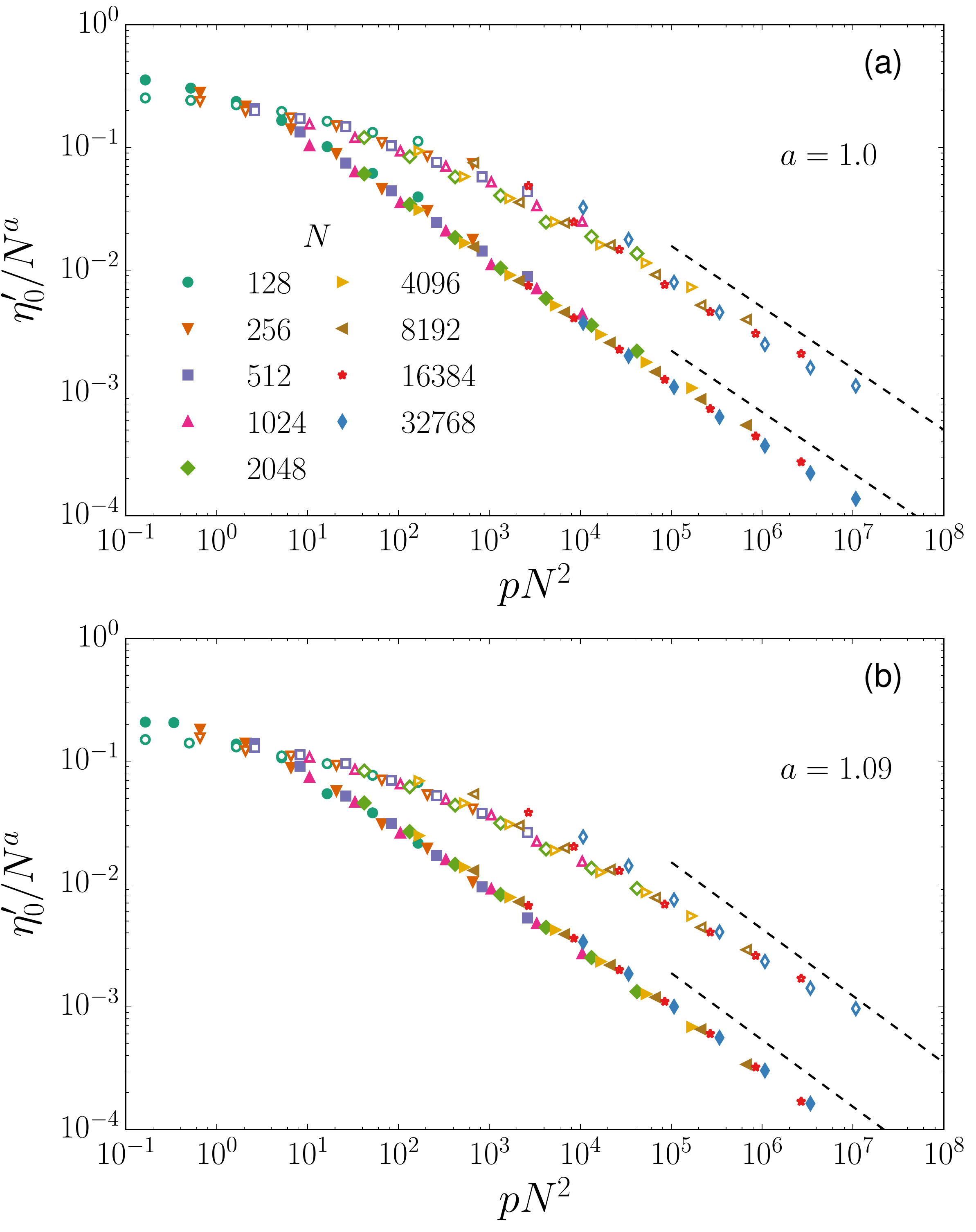}
 \caption{Finite size scaling collapse of the dynamic viscosity $\eta_0$ for \textcolor{black}{(a)} balanced contact damping and \textcolor{black}{(b)} Stokes drag. Filled/open data points are calculated with/without pre-stress. Dashed curves have a slope of $-a/2$, with $a$ indicated in the plot.}
 \label{fig:FS}
\end{figure}

Elastic moduli and the mean coordination number of marginally jammed matter are known to be influenced by finite size effects.\cite{tighe11b,goodrich12,dagois-bohy12,goodrich14,boschan16} In quasistatic systems they become important when the pressure $p$ is comparable to the pressure increment $p^* \sim 1/N^2$ required to add a contact to, or remove a contact from, the packing. Here we show that the same pressure scale governs  finite size effects in the dynamic viscosity $\eta_0$. 

In Fig.~\ref{fig:FS} the dynamic viscosities for both balanced contact damping and Stokes drag ($\eta_F = 1$)  are plotted for a wide range of pressures and system sizes, both with pre-stress (open symbols) and without pre-stress (filled symbols). In all cases, we find that the data collapse to a master curve when $\eta_0/N^a$ is plotted versus $p/p^* \sim pN^2$,  implying $\eta_0 \sim 1/p^{a/2}$.  For balanced contact damping, we find the best collapse when $a = 1.0$, consistent with the scaling $\eta_0 \sim 1/p^{1/2}$ determined above. For Stokes drag we find better collapse for the somewhat higher value $a = 1.09$. For comparison, we also plot curves with slope $-a/2$. Provided that $\eta_0$ is an intensive material property, as is typically the case, the master curves must approach this slope for large system sizes. This condition is met for the contact damping data, but for Stokes drag the collapsed data have a slightly shallower slope, particularly for the data with pre-stress. Using a lower value of $\alpha$ brings $-a/2$ closer to the observed slope, but the data collapse is somewhat worse. Given the small difference in these values and the scatter in our data, we consider it likely that $\alpha$ is in fact equal to 1 for Stokes drag. However, on the basis of present data we cannot exclude the possibility that $a >1$ for Stokes drag, or that $\eta_0$ has a weak system size dependence.

For both contact damping and Stokes drag, pre-stress plays a role in the onset of finite size effects. Whereas the data without pre-stress show a sharp crossover around $pN^2 \sim {\cal O}(1)$, the crossover in the data with pre-stress is much more gradual. Even for $pN^2 > 10^3$, a na\"ive power-law fit to $\eta_0$ versus $p$ would yield a slope that is too shallow. \textcolor{black}{Therefore studying the results of simulations with and without pre-stress, side-by-side, can potentially improve the assessment of critical exponents near jamming at modest system sizes.}

\section{Nonlinear Damping}

The drag forces considered in the previous sections are all linear in the particle velocities. Compared to nonlinear drag laws, linear forces are easier and cheaper to simulate. However, theory \cite{bretherton61,denkov08,seth11} and experiments \cite{katgert08} indicate that the bubble-bubble viscous force in foams (and so likely emulsions, as well) is in fact nonlinear in the relative velocity, as in Eq.~(\ref{eqn:nonlin}). We now probe the influence of an exponent $\alpha \neq 1$ on the complex shear modulus. Our main result is that the time scale $\tau_1$ must be generalized to account for a nontrivial frequency dependence. As a result, the frequency dependence of both the storage and the loss modulus changes.

Nonlinear equations of motion cannot be written as a matrix equation in terms of $\hat{\cal K}$ and $\hat{\cal B}$. Molecular dynamics simulations are an option,\cite{andreotti12} but beyond the scope of the present work. Instead, we turn to an approximation known as the method of equivalent damping.
The central idea of the approximation is to replace the nonlinear force law with an ``equivalent'' linear force law with a frequency dependent effective damping coefficient $k \tau_{\alpha}$, 
\begin{equation}
\vec f^{\, \rm eff}_{\alpha} = -k \tau_{\alpha} \, \Delta \vec v^{\,c} \,.
\label{eqn:eff}
\end{equation}
The effective damping coefficient is expressed in terms of a microscopic time scale $\tau_\alpha$ that depends on the frequency and amplitude of the forcing, as described below. $\tau_\alpha$ generalizes $\tau_1$, the constant time scale for $\alpha = 1$. 

We now apply the method of equivalent damping to a single degree of freedom system, namely an overdamped oscillator driven by a sinusoidal force with amplitude $F_0$ and frequency $\omega$. For the effective damping law of Eq.~(\ref{eqn:eff}), the resulting oscillations have an amplitude 
\begin{equation}
u_0 = \frac{F_0}{k} \left[ \frac{1}{ {1 + (\omega \tau_\alpha )^2}}\right]^{1/2} \,.
\end{equation}
To fix $\tau_\alpha$, we require that the energy dissipated by $\vec f^{\rm eff}_\alpha$ during one period is equal to the energy dissipated by the nonlinear force law (\ref{eqn:nonlin}) when the particle is constrained to follow the same trajectory through phase space. One finds
\begin{equation}
 \tau_{\alpha}(\omega) = 
\frac{2 \tau_1^\alpha}{\sqrt{\pi}} \left(\frac{u_0 \omega}{\rho_0}\right)^{\alpha - 1} \frac{\Gamma\left(1 + \frac{\alpha}{2}\right)}{\Gamma\left(\frac{3}{2} + \frac{\alpha}{2}\right)}
\,.
\end{equation}
This is an implicit relation, as $u_0$ depends on $\tau_\alpha$. Separately considering the low and high frequency limits gives
\begin{equation}
 \tau_{\alpha} = \left\{ 
   \begin{array}{cl} 
   1/(F_0 \omega)^{1-\alpha } & \,\,\, \omega < \omega_\times \\ 
   1/F_0^{(1-\alpha )/\alpha} &  \,\,\, \omega > \omega_\times\,,
   \end{array} \right. 
\label{eqn:taualpha}
\end{equation}
with a crossover frequency $\omega_\times \sim F_0^{(1-\alpha)/\alpha}$. 

To extend the above insights to soft sphere packings, we make an additional but reasonable assumption that the typical induced force on each contact is proportional to the applied stress, $F_0 \sim \delta \sigma $. 
Under this assumption, the scaling ansatz (\ref{eqn:ansatz}-\ref{eqn:calGpp}) remains valid, provided that one takes $\tau^* \sim \tau_\alpha/p$, instead of $\tau_1/p$. 
Because $\tau_\alpha$ is a function of frequency and the applied stress, the ``bare'' storage and loss moduli $G'$ and $G''$ (as opposed to ${\cal G}'$ and ${\cal G}''$) inherit new dependences on $\omega$ and $\delta \sigma$. 
For systems near jamming and the physically relevant case $\alpha < 1$,  $1/\tau^*$ is always smaller than $\omega_\times$ and hence $\tau_\alpha \sim (\delta \sigma \, \omega)^{\alpha - 1}$ in the quasistatic and critical regimes. In the quasistatic regime one finds that the storage modulus $G' \simeq G_0$ is unchanged, while the loss modulus becomes
\begin{equation}
G'' \sim  \frac{1}{\delta \sigma^{1-\alpha}} \frac{\omega^\alpha}{p^{1/2}} \,.
\label{eqn:Gpalpha}
\end{equation}
As in the linear case, the loss modulus in the quasistatic regime ``trivially'' reflects the form of the viscous force law, i.e.~both scale as $\omega^\alpha$. $G''$ also no longer displays linear response, as it depends on the applied stress.
In the critical regime one finds
\begin{equation}
G' \sim  \frac{\omega^{\alpha/2}}{\delta \sigma^{(1-\alpha)/2}}  
\end{equation}
and likewise
\begin{equation}
G'' \sim  \frac{\omega^{\alpha/2}}{\delta \sigma^{(1-\alpha)/2}}  \,.
\label{eqn:Gppalpha}
\end{equation}
We emphasize that the $\omega^{1/2}$ scaling of the linear case has been generalized to $\omega^{\alpha/2}$. Hence within the method of equivalent damping, the nonlinear frequency dependence of $G^*$ in viscous soft spheres contains a nontrivial dependence on the exponent $\alpha$ of the nonlinear viscous force law. 

\section{Conclusions}

We have shown that the viscoelastic response of viscous soft sphere packings close to jamming depends qualitatively on the damping law. The extent to which damping couples to floppy-like, and hence non-affine, motion is a key determinant of the resulting response. When the coupling is strong, as for balanced linear contact damping or Stokes-like drag with $\eta_F = 1$, the viscoelastic response displays dynamic critical scaling, including square root scaling of the storage and loss moduli over a broadening range of frequencies. When the coupling is weak, as when $0 < \beta < \beta^*$ for contact damping or when $\eta_F > 1/p^{1/2}$ for Stokes-like drag, aspects of the critical response are obscured. And when floppy-like motion is completely undamped, as for $\beta = 0$, dynamic critical scaling vanishes entirely. \textcolor{black}{We demonstrated a subtle interplay between the force law and non-affine correlations\textcolor{black}. For systems with contact damping, the only length scale identified by two-point correlation functions is the box size. However, in systems with Stokes drag, we observe a correlation length that diverges with vanishing $\omega$, with a cutoff at the box size. Finally, we presented numerical evidence that pre-stress increases the strength of finite size effects.}

We have also made predictions for the viscoelastic response in the presence of nonlinear drag laws. Within the context of the method of equivalent damping, we find that dynamic critical scaling survives; however the scaling of the bare storage and loss moduli now depends on the microscopic exponent $\alpha$. This provides a novel way to infer properties of the dominant dissipative mechanism at the particle scale from the frequency dependence of $G^*$. As the method of equivalent damping is an approximation, these predictions require further testing. As a basic check, we have verified Eq.~(\ref{eqn:Gpalpha}-\ref{eqn:Gppalpha}) by directly inserting the effective damping coefficient from Eq.~(\ref{eqn:taualpha}) in the linear equations of motion. Of course this does not constitute an independent test of the method of equivalent damping, which would require, e.g., molecular dynamics simulations of Durian's bubble model. We leave this as an important task for future work. 
 
Our results suggest that, when performing numerical studies of jammed matter, one must take care to match the form of the viscous force law to the physics of whatever particular material one wishes to model -- growing correlations do not wash out this detail. In particular, the linear contact damping law with $\beta = 0$ should be avoided, as it significantly alters the viscoelastic response and is difficult to justify on physical grounds, at least in the context of foams, emulsions, and soft colloidal particles.

\section{Acknowledgments}
This work was sponsored by NWO Exacte Wetenschappen (Physical Sciences) for the use of supercomputer facilities, with financial support from the Nederlandse Organisatie voor Wetenschappelijk Onderzoek (Netherlands Organization for Scientific Research, NWO).

%%%REFERENCES%%%
% \bibliography{tighe} %You need to replace "rsc" on this line with the name of your .bib file

\begin{mcitethebibliography}{59}
\providecommand*{\natexlab}[1]{#1}
\providecommand*{\mciteSetBstSublistMode}[1]{}
\providecommand*{\mciteSetBstMaxWidthForm}[2]{}
\providecommand*{\mciteBstWouldAddEndPuncttrue}
  {\def\EndOfBibitem{\unskip.}}
\providecommand*{\mciteBstWouldAddEndPunctfalse}
  {\let\EndOfBibitem\relax}
\providecommand*{\mciteSetBstMidEndSepPunct}[3]{}
\providecommand*{\mciteSetBstSublistLabelBeginEnd}[3]{}
\providecommand*{\EndOfBibitem}{}
\mciteSetBstSublistMode{f}
\mciteSetBstMaxWidthForm{subitem}
{(\emph{\alph{mcitesubitemcount}})}
\mciteSetBstSublistLabelBeginEnd{\mcitemaxwidthsubitemform\space}
{\relax}{\relax}

\bibitem[Durian(1995)]{durian95}
D.~J. Durian, \emph{Phys. Rev. Lett.}, 1995, \textbf{75}, 4780--4783\relax
\mciteBstWouldAddEndPuncttrue
\mciteSetBstMidEndSepPunct{\mcitedefaultmidpunct}
{\mcitedefaultendpunct}{\mcitedefaultseppunct}\relax
\EndOfBibitem
\bibitem[Tewari \emph{et~al.}(1999)Tewari, Schiemann, Durian, Knobler, Langer,
  and Liu]{tewari99}
S.~Tewari, D.~Schiemann, D.~J. Durian, C.~M. Knobler, S.~A. Langer and A.~J.
  Liu, \emph{Physical Review E}, 1999, \textbf{60}, 4385\relax
\mciteBstWouldAddEndPuncttrue
\mciteSetBstMidEndSepPunct{\mcitedefaultmidpunct}
{\mcitedefaultendpunct}{\mcitedefaultseppunct}\relax
\EndOfBibitem
\bibitem[Olsson and Teitel(2007)]{olsson07}
P.~Olsson and S.~Teitel, \emph{Phys.~Rev.~Lett.}, 2007, \textbf{99},
  178001\relax
\mciteBstWouldAddEndPuncttrue
\mciteSetBstMidEndSepPunct{\mcitedefaultmidpunct}
{\mcitedefaultendpunct}{\mcitedefaultseppunct}\relax
\EndOfBibitem
\bibitem[Langlois \emph{et~al.}(2008)Langlois, Hutzler, and Weaire]{langlois08}
V.~Langlois, S.~Hutzler and D.~Weaire, \emph{Phys.~Rev.~E}, 2008, \textbf{78},
  021401\relax
\mciteBstWouldAddEndPuncttrue
\mciteSetBstMidEndSepPunct{\mcitedefaultmidpunct}
{\mcitedefaultendpunct}{\mcitedefaultseppunct}\relax
\EndOfBibitem
\bibitem[Hatano(2009)]{hatano09}
T.~Hatano, \emph{Phys. Rev. E}, 2009, \textbf{79}, 050301\relax
\mciteBstWouldAddEndPuncttrue
\mciteSetBstMidEndSepPunct{\mcitedefaultmidpunct}
{\mcitedefaultendpunct}{\mcitedefaultseppunct}\relax
\EndOfBibitem
\bibitem[Tighe \emph{et~al.}(2010)Tighe, Woldhuis, Remmers, van Saarloos, and
  van Hecke]{tighe10c}
B.~P. Tighe, E.~Woldhuis, J.~J.~C. Remmers, W.~van Saarloos and M.~van Hecke,
  \emph{Phys. Rev. Lett.}, 2010, \textbf{105}, 088303\relax
\mciteBstWouldAddEndPuncttrue
\mciteSetBstMidEndSepPunct{\mcitedefaultmidpunct}
{\mcitedefaultendpunct}{\mcitedefaultseppunct}\relax
\EndOfBibitem
\bibitem[Boschan \emph{et~al.}(2016)Boschan, V{\aa}gberg, Somfai, and
  Tighe]{boschan16}
J.~Boschan, D.~V{\aa}gberg, E.~Somfai and B.~P. Tighe, \emph{Soft Matter},
  2016, \textbf{12}, 5450--5460\relax
\mciteBstWouldAddEndPuncttrue
\mciteSetBstMidEndSepPunct{\mcitedefaultmidpunct}
{\mcitedefaultendpunct}{\mcitedefaultseppunct}\relax
\EndOfBibitem
\bibitem[Boschan \emph{et~al.}(2017)Boschan, Vasudavan, Boukany, Somfai, and
  Tighe]{boschan17}
J.~Boschan, S.~Vasudavan, P.~Boukany, E.~Somfai and B.~Tighe, \emph{Soft
  Matter}, 2017,  DOI: 10.1039/C7SM01700F\relax
\mciteBstWouldAddEndPuncttrue
\mciteSetBstMidEndSepPunct{\mcitedefaultmidpunct}
{\mcitedefaultendpunct}{\mcitedefaultseppunct}\relax
\EndOfBibitem
\bibitem[O'Hern \emph{et~al.}(2003)O'Hern, Silbert, Liu, and Nagel]{ohern03}
C.~S. O'Hern, L.~E. Silbert, A.~J. Liu and S.~R. Nagel, \emph{Phys.~Rev.~E},
  2003, \textbf{68}, 011306\relax
\mciteBstWouldAddEndPuncttrue
\mciteSetBstMidEndSepPunct{\mcitedefaultmidpunct}
{\mcitedefaultendpunct}{\mcitedefaultseppunct}\relax
\EndOfBibitem
\bibitem[Bolton and Weaire(1990)]{bolton90}
F.~Bolton and D.~Weaire, \emph{Phys. Rev. Lett.}, 1990, \textbf{65},
  3449--3451\relax
\mciteBstWouldAddEndPuncttrue
\mciteSetBstMidEndSepPunct{\mcitedefaultmidpunct}
{\mcitedefaultendpunct}{\mcitedefaultseppunct}\relax
\EndOfBibitem
\bibitem[Silbert \emph{et~al.}(2005)Silbert, Liu, and Nagel]{silbert05}
L.~E. Silbert, A.~J. Liu and S.~R. Nagel, \emph{Phys.~Rev.~Lett.}, 2005,
  \textbf{95}, 098301\relax
\mciteBstWouldAddEndPuncttrue
\mciteSetBstMidEndSepPunct{\mcitedefaultmidpunct}
{\mcitedefaultendpunct}{\mcitedefaultseppunct}\relax
\EndOfBibitem
\bibitem[Ellenbroek \emph{et~al.}(2006)Ellenbroek, Somfai, van Hecke, and van
  Saarloos]{ellenbroek06}
W.~G. Ellenbroek, E.~Somfai, M.~van Hecke and W.~van Saarloos,
  \emph{Phys.~Rev.~Lett.}, 2006, \textbf{97}, 258001\relax
\mciteBstWouldAddEndPuncttrue
\mciteSetBstMidEndSepPunct{\mcitedefaultmidpunct}
{\mcitedefaultendpunct}{\mcitedefaultseppunct}\relax
\EndOfBibitem
\bibitem[Tighe(2011)]{tighe11}
B.~P. Tighe, \emph{Phys. Rev. Lett.}, 2011, \textbf{107}, 158303\relax
\mciteBstWouldAddEndPuncttrue
\mciteSetBstMidEndSepPunct{\mcitedefaultmidpunct}
{\mcitedefaultendpunct}{\mcitedefaultseppunct}\relax
\EndOfBibitem
\bibitem[Lerner \emph{et~al.}(2014)Lerner, DeGiuli, D{\"u}ring, and
  Wyart]{lerner14}
E.~Lerner, E.~DeGiuli, G.~D{\"u}ring and M.~Wyart, \emph{Soft Matter}, 2014,
  \textbf{10}, 5085--5092\relax
\mciteBstWouldAddEndPuncttrue
\mciteSetBstMidEndSepPunct{\mcitedefaultmidpunct}
{\mcitedefaultendpunct}{\mcitedefaultseppunct}\relax
\EndOfBibitem
\bibitem[Karimi and Maloney(2015)]{karimi15}
K.~Karimi and C.~E. Maloney, \emph{Phys. Rev. E}, 2015, \textbf{92},
  022208\relax
\mciteBstWouldAddEndPuncttrue
\mciteSetBstMidEndSepPunct{\mcitedefaultmidpunct}
{\mcitedefaultendpunct}{\mcitedefaultseppunct}\relax
\EndOfBibitem
\bibitem[Baumgarten \emph{et~al.}(2017)Baumgarten, V{\aa}gberg, and
  Tighe]{baumgarten17}
K.~Baumgarten, D.~V{\aa}gberg and B.~P. Tighe, \emph{Phys. Rev. Lett.}, 2017,
  \textbf{118}, 098001\relax
\mciteBstWouldAddEndPuncttrue
\mciteSetBstMidEndSepPunct{\mcitedefaultmidpunct}
{\mcitedefaultendpunct}{\mcitedefaultseppunct}\relax
\EndOfBibitem
\bibitem[Khakalo \emph{et~al.}(2017)Khakalo, Baumgarten, Tighe, and
  Puisto]{khakalo17}
K.~Khakalo, K.~Baumgarten, B.~P. Tighe and A.~Puisto, \emph{arXiv:1706.03932},
  2017\relax
\mciteBstWouldAddEndPuncttrue
\mciteSetBstMidEndSepPunct{\mcitedefaultmidpunct}
{\mcitedefaultendpunct}{\mcitedefaultseppunct}\relax
\EndOfBibitem
\bibitem[Barnes and Hutton(1989)]{barnes}
H.~A. Barnes and J.~F. Hutton, \emph{An Introduction to Rheology}, Elsevier,
  1989\relax
\mciteBstWouldAddEndPuncttrue
\mciteSetBstMidEndSepPunct{\mcitedefaultmidpunct}
{\mcitedefaultendpunct}{\mcitedefaultseppunct}\relax
\EndOfBibitem
\bibitem[Liu and Nagel(1998)]{liu98}
A.~J. Liu and S.~R. Nagel, \emph{Nature}, 1998, \textbf{396}, 21--22\relax
\mciteBstWouldAddEndPuncttrue
\mciteSetBstMidEndSepPunct{\mcitedefaultmidpunct}
{\mcitedefaultendpunct}{\mcitedefaultseppunct}\relax
\EndOfBibitem
\bibitem[Cohen-Addad \emph{et~al.}(1998)Cohen-Addad, Hoballah, and
  H\"ohler]{cohenaddad98}
S.~Cohen-Addad, H.~Hoballah and R.~H\"ohler, \emph{Phys. Rev. E}, 1998,
  \textbf{57}, 6897--6901\relax
\mciteBstWouldAddEndPuncttrue
\mciteSetBstMidEndSepPunct{\mcitedefaultmidpunct}
{\mcitedefaultendpunct}{\mcitedefaultseppunct}\relax
\EndOfBibitem
\bibitem[Gopal and Durian(2003)]{gopal03}
A.~D. Gopal and D.~J. Durian, \emph{Phys.~Rev.~Lett.}, 2003, \textbf{91},
  188303\relax
\mciteBstWouldAddEndPuncttrue
\mciteSetBstMidEndSepPunct{\mcitedefaultmidpunct}
{\mcitedefaultendpunct}{\mcitedefaultseppunct}\relax
\EndOfBibitem
\bibitem[Kropka and Celina(2010)]{kropka10}
J.~M. Kropka and M.~Celina, \emph{J. Chem. Phys.}, 2010, \textbf{133},
  024904\relax
\mciteBstWouldAddEndPuncttrue
\mciteSetBstMidEndSepPunct{\mcitedefaultmidpunct}
{\mcitedefaultendpunct}{\mcitedefaultseppunct}\relax
\EndOfBibitem
\bibitem[Li{\'e}tor-Santos \emph{et~al.}(2011)Li{\'e}tor-Santos,
  Sierra-Mart{\'\i}n, and Fern{\'a}ndez-Nieves]{lietorsantos11}
J.~J. Li{\'e}tor-Santos, B.~Sierra-Mart{\'\i}n and A.~Fern{\'a}ndez-Nieves,
  \emph{Phys. Rev. E}, 2011, \textbf{84}, 060402\relax
\mciteBstWouldAddEndPuncttrue
\mciteSetBstMidEndSepPunct{\mcitedefaultmidpunct}
{\mcitedefaultendpunct}{\mcitedefaultseppunct}\relax
\EndOfBibitem
\bibitem[Srivastava and Fisher(2017)]{srivastava17}
I.~Srivastava and T.~S. Fisher, \emph{Soft Matter}, 2017, \textbf{13},
  3411--3421\relax
\mciteBstWouldAddEndPuncttrue
\mciteSetBstMidEndSepPunct{\mcitedefaultmidpunct}
{\mcitedefaultendpunct}{\mcitedefaultseppunct}\relax
\EndOfBibitem
\bibitem[Bretherton(1961)]{bretherton61}
F.~P. Bretherton, \emph{J. Fluid Mech.}, 1961, \textbf{10}, 166\relax
\mciteBstWouldAddEndPuncttrue
\mciteSetBstMidEndSepPunct{\mcitedefaultmidpunct}
{\mcitedefaultendpunct}{\mcitedefaultseppunct}\relax
\EndOfBibitem
\bibitem[Denkov \emph{et~al.}(2008)Denkov, Tcholakova, Golemanov,
  Ananthapadmanabhan, and Lips]{denkov08}
N.~Denkov, S.~Tcholakova, K.~Golemanov, K.~Ananthapadmanabhan and A.~Lips,
  \emph{Phys. Rev. Lett.}, 2008, \textbf{100}, 138301\relax
\mciteBstWouldAddEndPuncttrue
\mciteSetBstMidEndSepPunct{\mcitedefaultmidpunct}
{\mcitedefaultendpunct}{\mcitedefaultseppunct}\relax
\EndOfBibitem
\bibitem[Seth \emph{et~al.}(2011)Seth, Mohan, Locatelli-Champagne, Cloitre, and
  Bonnecaze]{seth11}
J.~R. Seth, L.~Mohan, C.~Locatelli-Champagne, M.~Cloitre and R.~T. Bonnecaze,
  \emph{Nat Mater}, 2011, \textbf{10}, 838--843\relax
\mciteBstWouldAddEndPuncttrue
\mciteSetBstMidEndSepPunct{\mcitedefaultmidpunct}
{\mcitedefaultendpunct}{\mcitedefaultseppunct}\relax
\EndOfBibitem
\bibitem[Andreotti \emph{et~al.}(2012)Andreotti, Barrat, and
  Heussinger]{andreotti12}
B.~Andreotti, J.-L. Barrat and C.~Heussinger, \emph{Phys. Rev. Lett.}, 2012,
  \textbf{109}, 105901\relax
\mciteBstWouldAddEndPuncttrue
\mciteSetBstMidEndSepPunct{\mcitedefaultmidpunct}
{\mcitedefaultendpunct}{\mcitedefaultseppunct}\relax
\EndOfBibitem
\bibitem[Halperin and Hohenberg(1969)]{hohenberghalperin}
B.~I. Halperin and P.~C. Hohenberg, \emph{Phys. Rev.}, 1969, \textbf{177},
  952--971\relax
\mciteBstWouldAddEndPuncttrue
\mciteSetBstMidEndSepPunct{\mcitedefaultmidpunct}
{\mcitedefaultendpunct}{\mcitedefaultseppunct}\relax
\EndOfBibitem
\bibitem[Koeze \emph{et~al.}(2016)Koeze, V{\aa}gberg, Tjoa, and Tighe]{koeze16}
D.~J. Koeze, D.~V{\aa}gberg, B.~B. Tjoa and B.~P. Tighe, \emph{EPL}, 2016,
  \textbf{113}, 54001\relax
\mciteBstWouldAddEndPuncttrue
\mciteSetBstMidEndSepPunct{\mcitedefaultmidpunct}
{\mcitedefaultendpunct}{\mcitedefaultseppunct}\relax
\EndOfBibitem
\bibitem[Goodrich \emph{et~al.}(2012)Goodrich, Liu, and Nagel]{goodrich12}
C.~P. Goodrich, A.~J. Liu and S.~R. Nagel, \emph{Phys. Rev. Lett.}, 2012,
  \textbf{109}, 095704\relax
\mciteBstWouldAddEndPuncttrue
\mciteSetBstMidEndSepPunct{\mcitedefaultmidpunct}
{\mcitedefaultendpunct}{\mcitedefaultseppunct}\relax
\EndOfBibitem
\bibitem[van Hecke(2010)]{vanhecke10}
M.~van Hecke, \emph{J.~Phys.~Cond.~Matt.}, 2010, \textbf{22}, 033101\relax
\mciteBstWouldAddEndPuncttrue
\mciteSetBstMidEndSepPunct{\mcitedefaultmidpunct}
{\mcitedefaultendpunct}{\mcitedefaultseppunct}\relax
\EndOfBibitem
\bibitem[Liu and Nagel(2010)]{liu10a}
A.~J. Liu and S.~R. Nagel, \emph{Ann. Rev. Cond. Matt. Phys.}, 2010,
  \textbf{1}, 347--369\relax
\mciteBstWouldAddEndPuncttrue
\mciteSetBstMidEndSepPunct{\mcitedefaultmidpunct}
{\mcitedefaultendpunct}{\mcitedefaultseppunct}\relax
\EndOfBibitem
\bibitem[Morse and Witten(1993)]{morse93}
D.~Morse and T.~Witten, \emph{EPL}, 1993, \textbf{22}, 549\relax
\mciteBstWouldAddEndPuncttrue
\mciteSetBstMidEndSepPunct{\mcitedefaultmidpunct}
{\mcitedefaultendpunct}{\mcitedefaultseppunct}\relax
\EndOfBibitem
\bibitem[Lacasse \emph{et~al.}(1996)Lacasse, Grest, and Levine]{lacasse96}
M.-D. Lacasse, G.~S. Grest and D.~Levine, \emph{Phys. Rev. E}, 1996,
  \textbf{54}, 5436\relax
\mciteBstWouldAddEndPuncttrue
\mciteSetBstMidEndSepPunct{\mcitedefaultmidpunct}
{\mcitedefaultendpunct}{\mcitedefaultseppunct}\relax
\EndOfBibitem
\bibitem[Tighe(2016)]{tighechapter}
B.~P. Tighe, in \emph{Handbook of Granular Materials}, ed. S.~V. Franklin and
  M.~D. Shattuck, CRC Press, 2016, ch. Wet Foams, Slippery Grains\relax
\mciteBstWouldAddEndPuncttrue
\mciteSetBstMidEndSepPunct{\mcitedefaultmidpunct}
{\mcitedefaultendpunct}{\mcitedefaultseppunct}\relax
\EndOfBibitem
\bibitem[Hutzler \emph{et~al.}(2014)Hutzler, Murtagh, Whyte, Tobin, and
  Weaire]{hutzler14}
S.~Hutzler, R.~P. Murtagh, D.~Whyte, S.~T. Tobin and D.~Weaire, \emph{Soft
  Matter}, 2014, \textbf{10}, 7103--7108\relax
\mciteBstWouldAddEndPuncttrue
\mciteSetBstMidEndSepPunct{\mcitedefaultmidpunct}
{\mcitedefaultendpunct}{\mcitedefaultseppunct}\relax
\EndOfBibitem
\bibitem[H{\"o}hler and Cohen-Addad(2017)]{hohler17}
R.~H{\"o}hler and S.~Cohen-Addad, \emph{Soft matter}, 2017, \textbf{13},
  1371--1383\relax
\mciteBstWouldAddEndPuncttrue
\mciteSetBstMidEndSepPunct{\mcitedefaultmidpunct}
{\mcitedefaultendpunct}{\mcitedefaultseppunct}\relax
\EndOfBibitem
\bibitem[Winkelmann \emph{et~al.}(2017)Winkelmann, Dunne, Langlois, M{\"o}bius,
  Weaire, and Hutzler]{winkelmann17}
J.~Winkelmann, F.~Dunne, V.~Langlois, M.~M{\"o}bius, D.~Weaire and S.~Hutzler,
  \emph{Colloids and Surfaces A: Physicochemical and Engineering Aspects},
  2017\relax
\mciteBstWouldAddEndPuncttrue
\mciteSetBstMidEndSepPunct{\mcitedefaultmidpunct}
{\mcitedefaultendpunct}{\mcitedefaultseppunct}\relax
\EndOfBibitem
\bibitem[Maloney and Robbins(2008)]{maloney08}
C.~E. Maloney and M.~O. Robbins, \emph{Journal of Physics: Condensed Matter},
  2008, \textbf{20}, 244128\relax
\mciteBstWouldAddEndPuncttrue
\mciteSetBstMidEndSepPunct{\mcitedefaultmidpunct}
{\mcitedefaultendpunct}{\mcitedefaultseppunct}\relax
\EndOfBibitem
\bibitem[Vagberg and Tighe(2017)]{vagberg17}
D.~Vagberg and B.~P. Tighe, \emph{arXiv:1706.06378}, 2017\relax
\mciteBstWouldAddEndPuncttrue
\mciteSetBstMidEndSepPunct{\mcitedefaultmidpunct}
{\mcitedefaultendpunct}{\mcitedefaultseppunct}\relax
\EndOfBibitem
\bibitem[Peyneau and Roux(2008)]{peyneau08}
P.-E. Peyneau and J.-N. Roux, \emph{Phys. Rev. E}, 2008, \textbf{78},
  011307\relax
\mciteBstWouldAddEndPuncttrue
\mciteSetBstMidEndSepPunct{\mcitedefaultmidpunct}
{\mcitedefaultendpunct}{\mcitedefaultseppunct}\relax
\EndOfBibitem
\bibitem[Hatano(2008)]{hatano08}
T.~Hatano, \emph{J.PHYS.SOC.JPN.}, 2008, \textbf{77}, 123002\relax
\mciteBstWouldAddEndPuncttrue
\mciteSetBstMidEndSepPunct{\mcitedefaultmidpunct}
{\mcitedefaultendpunct}{\mcitedefaultseppunct}\relax
\EndOfBibitem
\bibitem[Otsuki and Hayakawa(2009)]{otsuki09}
M.~Otsuki and H.~Hayakawa, \emph{Phys. Rev. E}, 2009, \textbf{80}, 011308\relax
\mciteBstWouldAddEndPuncttrue
\mciteSetBstMidEndSepPunct{\mcitedefaultmidpunct}
{\mcitedefaultendpunct}{\mcitedefaultseppunct}\relax
\EndOfBibitem
\bibitem[Heussinger \emph{et~al.}(2010)Heussinger, Chaudhuri, and
  Barrat]{heussinger10}
C.~Heussinger, P.~Chaudhuri and J.-L. Barrat, \emph{Soft Matter}, 2010,
  \textbf{6}, year\relax
\mciteBstWouldAddEndPuncttrue
\mciteSetBstMidEndSepPunct{\mcitedefaultmidpunct}
{\mcitedefaultendpunct}{\mcitedefaultseppunct}\relax
\EndOfBibitem
\bibitem[Evans and Morriss(2008)]{evansmorriss}
D.~J. Evans and G.~Morriss, \emph{Statistical Mechanics of Nonequilibrium
  Liquids}, Cambridge University Press, 2008\relax
\mciteBstWouldAddEndPuncttrue
\mciteSetBstMidEndSepPunct{\mcitedefaultmidpunct}
{\mcitedefaultendpunct}{\mcitedefaultseppunct}\relax
\EndOfBibitem
\bibitem[Ikeda \emph{et~al.}(2012)Ikeda, Berthier, and Sollich]{ikeda12}
A.~Ikeda, L.~Berthier and P.~Sollich, \emph{Phys. Rev. Lett.}, 2012,
  \textbf{109}, 018301\relax
\mciteBstWouldAddEndPuncttrue
\mciteSetBstMidEndSepPunct{\mcitedefaultmidpunct}
{\mcitedefaultendpunct}{\mcitedefaultseppunct}\relax
\EndOfBibitem
\bibitem[Tighe(2012)]{tighe12}
B.~P. Tighe, \emph{Phys. Rev. Lett.}, 2012, \textbf{109}, 168303\relax
\mciteBstWouldAddEndPuncttrue
\mciteSetBstMidEndSepPunct{\mcitedefaultmidpunct}
{\mcitedefaultendpunct}{\mcitedefaultseppunct}\relax
\EndOfBibitem
\bibitem[Lerner \emph{et~al.}(2012)Lerner, D\"uring, and Wyart]{lerner12}
E.~Lerner, G.~D\"uring and M.~Wyart, \emph{Proc. Nat. Acad. Sci.}, 2012,
  \textbf{109}, 4798--4803\relax
\mciteBstWouldAddEndPuncttrue
\mciteSetBstMidEndSepPunct{\mcitedefaultmidpunct}
{\mcitedefaultendpunct}{\mcitedefaultseppunct}\relax
\EndOfBibitem
\bibitem[Katgert \emph{et~al.}(2008)Katgert, M\"obius, and van
  Hecke]{katgert08}
G.~Katgert, M.~E. M\"obius and M.~van Hecke, \emph{Phys.~Rev.~Lett.}, 2008,
  \textbf{101}, 058301\relax
\mciteBstWouldAddEndPuncttrue
\mciteSetBstMidEndSepPunct{\mcitedefaultmidpunct}
{\mcitedefaultendpunct}{\mcitedefaultseppunct}\relax
\EndOfBibitem
\bibitem[Ellenbroek \emph{et~al.}(2009)Ellenbroek, van Hecke, and van
  Saarloos]{ellenbroek09}
W.~G. Ellenbroek, M.~van Hecke and W.~van Saarloos, \emph{Phys. Rev. E}, 2009,
  \textbf{80}, 061307\relax
\mciteBstWouldAddEndPuncttrue
\mciteSetBstMidEndSepPunct{\mcitedefaultmidpunct}
{\mcitedefaultendpunct}{\mcitedefaultseppunct}\relax
\EndOfBibitem
\bibitem[Wyart(2005)]{wyartannales}
M.~Wyart, \emph{Annales de Physique}, 2005, \textbf{30}, 1\relax
\mciteBstWouldAddEndPuncttrue
\mciteSetBstMidEndSepPunct{\mcitedefaultmidpunct}
{\mcitedefaultendpunct}{\mcitedefaultseppunct}\relax
\EndOfBibitem
\bibitem[Dagois-Bohy \emph{et~al.}(2012)Dagois-Bohy, Tighe, Simon, Henkes, and
  van Hecke]{dagois-bohy12}
S.~Dagois-Bohy, B.~P. Tighe, J.~Simon, S.~Henkes and M.~van Hecke, \emph{Phys.
  Rev. Lett.}, 2012, \textbf{109}, 095703\relax
\mciteBstWouldAddEndPuncttrue
\mciteSetBstMidEndSepPunct{\mcitedefaultmidpunct}
{\mcitedefaultendpunct}{\mcitedefaultseppunct}\relax
\EndOfBibitem
\bibitem[Wyart \emph{et~al.}(2008)Wyart, Liang, Kabla, and Mahadevan]{wyart08}
M.~Wyart, H.~Liang, A.~Kabla and L.~Mahadevan, \emph{Phys. Rev. Lett.}, 2008,
  \textbf{101}, 215501\relax
\mciteBstWouldAddEndPuncttrue
\mciteSetBstMidEndSepPunct{\mcitedefaultmidpunct}
{\mcitedefaultendpunct}{\mcitedefaultseppunct}\relax
\EndOfBibitem
\bibitem[DiDonna and Lubensky(2005)]{didonna05}
B.~A. DiDonna and T.~C. Lubensky, \emph{Phys.~Rev.~E}, 2005, \textbf{72},
  066619\relax
\mciteBstWouldAddEndPuncttrue
\mciteSetBstMidEndSepPunct{\mcitedefaultmidpunct}
{\mcitedefaultendpunct}{\mcitedefaultseppunct}\relax
\EndOfBibitem
\bibitem[Maloney and Lema\^\i{}tre(2006)]{maloney06}
C.~E. Maloney and A.~Lema\^\i{}tre, \emph{Phys. Rev. E}, 2006, \textbf{74},
  016118\relax
\mciteBstWouldAddEndPuncttrue
\mciteSetBstMidEndSepPunct{\mcitedefaultmidpunct}
{\mcitedefaultendpunct}{\mcitedefaultseppunct}\relax
\EndOfBibitem
\bibitem[Tighe \emph{et~al.}(2010)Tighe, Snoeijer, Vlugt, and van
  Hecke]{tighe10}
B.~P. Tighe, J.~H. Snoeijer, T.~J.~H. Vlugt and M.~van Hecke, \emph{Soft
  Matter}, 2010, \textbf{6}, 2908\relax
\mciteBstWouldAddEndPuncttrue
\mciteSetBstMidEndSepPunct{\mcitedefaultmidpunct}
{\mcitedefaultendpunct}{\mcitedefaultseppunct}\relax
\EndOfBibitem
\bibitem[Goodrich \emph{et~al.}(2014)Goodrich, Dagois-Bohy, Tighe, van Hecke,
  Liu, and Nagel]{goodrich14}
C.~P. Goodrich, S.~Dagois-Bohy, B.~P. Tighe, M.~van Hecke, A.~J. Liu and S.~R.
  Nagel, \emph{Phys. Rev. E}, 2014, \textbf{90}, 022138\relax
\mciteBstWouldAddEndPuncttrue
\mciteSetBstMidEndSepPunct{\mcitedefaultmidpunct}
{\mcitedefaultendpunct}{\mcitedefaultseppunct}\relax
\EndOfBibitem
\bibitem[Tighe and Vlugt(2011)]{tighe11b}
B.~P. Tighe and T.~J.~H. Vlugt, \emph{Journal of Statistical Mechanics: Theory
  and Experiment}, 2011,  P04002\relax
\mciteBstWouldAddEndPuncttrue
\mciteSetBstMidEndSepPunct{\mcitedefaultmidpunct}
{\mcitedefaultendpunct}{\mcitedefaultseppunct}\relax
\EndOfBibitem
\end{mcitethebibliography}
% \bibliographystyle{rsc} %the RSC's .bst file

\providecommand*{\mcitethebibliography}{\thebibliography}
\csname @ifundefined\endcsname{endmcitethebibliography}
{\let\endmcitethebibliography\endthebibliography}{}

\end{document}